\documentclass[pra,floatfix,twocolumn,showpacs,superscriptaddress]{revtex4}
\pdfoutput=1
\usepackage{graphicx}
\usepackage{amssymb}
\usepackage{amsmath,amsfonts}
\usepackage{epstopdf}

\newcommand {\ket}[1] {|#1 \rangle}
\newcommand {\bra}[1] {\langle#1 |}

\newcommand {\ketbra}[2] {| #1 \rangle \langle #2 |}

\newcommand {\av}[1] {\langle #1 \rangle}

\begin{document}

\title{Collective spin systems in dispersive optical cavity QED: Quantum phase transitions and entanglement}

\author{S. Morrison}
\affiliation{Institute for Theoretical Physics, University of Innsbruck, Innsbruck A-6020, Austria}
\affiliation{Institute for Quantum Optics and Quantum Information of the Austrian Academy of
Sciences, Innsbruck A-6020, Austria}
\affiliation{Department of Physics, University of Auckland, Private Bag 92019, Auckland, New Zealand}

\author{A.~S. Parkins}
\affiliation{Department of Physics, University of Auckland, Private Bag 92019, Auckland, New Zealand}

\date{\today}

\begin{abstract}
We propose a cavity QED setup which implements a dissipative Lipkin-Meshkov-Glick model -- an interacting collective spin system. By varying the external model parameters the system can be made to undergo both first-and second-order quantum phase transitions, which are signified by dramatic changes in cavity output field properties, such as the probe laser transmission spectrum. The steady-state entanglement between pairs of atoms is shown to peak at the critical points and can be experimentally determined by suitable measurements on the cavity output field. The entanglement dynamics also exhibits pronounced variations in the vicinities of the phase transitions.
\end{abstract}

\maketitle


\section{Introduction} \label{sect:introduction}

The branch of atomic physics associated with ultracold atoms, ions, and molecules now provides a rich and exciting arena for investigations of strongly interacting, many-body quantum systems. Trapping and cooling techniques, coherent laser or microwave interactions, and applied magnetic fields enable exquisite control of both external (motional) and internal (electronic) degrees of freedom of the particles, allowing one to ``tailor'' particle-particle interactions and thereby implement a broad range of systems that can be described accurately and transparently by idealized (but nontrivial) many-body Hamiltonians. An important example is the Hubbard model, realized with ultracold atoms in periodic optical lattices~\cite{Jaksch05,Greiner02}, while realizations of other novel and significant lattice-spin models have been proposed, for example, with dipolar molecules in optical lattices~\cite{Micheli06} and with chains of trapped atomic ions~\cite{Porras04}.
The common, defining feature of these systems is the possibility for quantum critical phenomena, i.e., transitions between distinct quantum phases, in response to variations of an effective field or particle-particle interaction strength around some critical value.

The above-mentioned schemes generally provide many-body quantum systems that are subject to short-range (e.g., nearest-neighbor) interactions. Another interesting and commonly studied class of many-body systems are those possessing long-range, or even infinite-range, interactions, for which theoretical models typically allow exact solutions in the thermodynamic limit, or at least enable efficient numerical solution for large numbers of particles.
A standard and classic example is the Lipkin-Meshkov-Glick (LMG) model~\cite{originalLMG123}, which was originally introduced in nuclear physics and is described by a Hamiltonian of the form
\begin{equation} \label{LMGHamiltonian}
H_\textrm{LMG} =  -2 h J_z  - \frac{2\lambda}{N} (J_x^2 +\gamma J_y^2),
\end{equation}
where $\{ J_x,J_y,J_z\}$ are collective angular momentum operators for $N$ spin-1/2 particles, $h$ and $\lambda$ are parameters giving the effective magnetic field and spin-spin interaction strengths, respectively, and $\gamma\in[-1,1]$ is an anisotropy parameter.
In this model, each spin interacts identically with every other spin and the nature of this interaction may be ferromagnetic ($\lambda>0$) or antiferromagnetic ($\lambda<0$). Significantly, the model exhibits critical behavior at zero temperature; in particular, either first- or second-order quantum phase transitions may occur (depending on the choice of $\lambda$ and $\gamma$) as the ratio between $\lambda$ and $h$ is varied across a critical value.

This quantum critical behavior, combined with the relative simplicity of the model, has led to renewed theoretical interest in the LMG model from the point of view of studying entanglement properties of many-particle systems in relation to quantum phase transitions~\cite{Osterloh02,Osborne02,GVidal03}. Bipartite entanglement measures characterizing entanglement between a pair of spins (e.g., the concurrence) or between two blocks of spins (e.g., the entanglement entropy) are relatively straightforward to compute for the LMG model and can display marked critical behavior and scaling at quantum critical points~\cite{EntLMGSecondOrder,EntLMGFirstOrder,EntLMGDynamics,EntLMGEntropy,EntLMGCUT,EntLMGBlock,EntLMGConcurrReview}.

Given these interesting and very topical features of the LMG model, it follows that the physical realization of a system described accurately by such a model would provide a valuable test bed for studies of quantum critical phenomena and entanglement. However, the question naturally arises as to how realistic such an idealized model could be; the assumption of ``infinite-range'' interactions is obviously demanding and implies a very specialized system. Hamiltonians of the form~(\ref{LMGHamiltonian}) (with $\gamma=0$) have appeared recently in reduced two-mode models of atomic Bose-Einstein condensates undergoing tunnelling in double-well potentials or transitions between two internal atomic states~\cite{BECLMGSemiclassical,BECMicheli}, and in models of a few trapped ions interacting with laser fields~\cite{MolmerGHZ,Fleischhauer1}, but emphasis in these works has been on unitary or adiabatic evolution from some initial atomic state to some final, prescribed (entangled) state, while flexibility of these systems with respect to parameters of the LMG model (i.e., $\lambda$, $N$, $\gamma$) appears limited.

Another possibility, furnished by the field of quantum optics, and for which long-range atom-atom interactions actually occur quite naturally, is cavity quantum electrodynamics (cavity QED)~\cite{Berman94}. Here, one considers ensembles of atoms interacting, through an electronic transition, with a common electromagnetic field mode supported by an optical resonator. Through this common coupling, the field mode can effectively mediate interactions between atoms located at quite arbitrary and separate positions within the mode. So, in particular, the concept of an interaction ``length'' becomes redundant in this setting and a collective description of the atoms is appropriate.

In fact, that an ensemble of atoms coupled to a common field mode can be viewed as a many-body system of interacting spins was highlighted many years ago with the prediction of a thermal equilibrium phase transition in the celebrated Dicke model of $N$ two-level atoms coupled to a single quantized field mode~\cite{Hepp73,Wang73,Hioe73,Carmichael73,Duncan74},
\begin{eqnarray}\label{DickeHamiltonian}
H = \omega a^\dag a + \omega_0 J_z + \frac{\lambda}{\sqrt{N}} (a^\dag + a)(J_+ + J_-) ,
\end{eqnarray}
where $a$ is the annihilation operator for the field mode of frequency $\omega$, $\omega_0$ is the atomic transition frequency, and $\lambda$ is the atom-field coupling strength  (we set $\hbar=1$).
In particular, above a certain critical value of the coupling strength the system enters a so-called ``superradiant'' phase~\cite{Dicke}.
This phase transition persists at zero temperature~\cite{Emary03a,Emary03b}, with associated critical behavior of both the atom-field and atom-atom quantum entanglement
\cite{Lambert04,Lambert05,Reslen05}. The critical coupling strength at zero temperature is given by $\lambda_{\rm c}=\sqrt{\omega\omega_0}/2$, which means that $\lambda$ must be comparable to the field and/or atomic transition frequencies if the transition regime is to be reached. For atomic dipole transitions, this is typically not the case and, in fact, if it happened to be so, then the model~(\ref{DickeHamiltonian}) would be inadequate; in particular, the $A^2$ term [omitted from~(\ref{DickeHamiltonian})] of the minimal coupling Hamiltonian should be included and doing so one actually finds that no phase transition exists~\cite{Rzaznewski75}.

However, a recent proposal for realizing the Dicke model quantum phase transition, based on Raman transitions between stable atomic ground states in an optical cavity QED setting~\cite{Dimer07}, circumvents these issues by (i) implementing a system in which the relevant frequency and coupling scales are determined by light-induced frequency shifts and Raman transition rates, and (ii) utilizing an open-system dynamics (as opposed to a closed, Hamiltonian system) with input and output fields (i.e., external laser fields and cavity mode losses), thereby replacing a (fragile) thermal equilibrium phase transition with a (robust) dynamical, nonequilibrium phase transition.
Furthermore, as shown in Ref.~\cite{Dimer07}, the cavity output field offers a unique window on the system's behavior and properties, with, for example, fluorescence and quadrature-variance measurements providing dramatic signatures of criticality in the system, as well as quantitative measures of fluctuations and entanglement.

These features of the optical cavity QED system, combined with the observation that, in the dispersive limit $\omega\gg\{\omega_0,\lambda\}$, the cavity mode may be adiabatically eliminated and the Dicke Hamiltonian reduced to the form
\begin{eqnarray}
H = \omega_0 J_z - \frac{4\lambda^2}{N\omega}\, J_x^2 ,
\end{eqnarray}
where $J_x=\frac{1}{2}(J_++J_-)$, motivate us to explore the possibilities for studying the LMG model in such a setting. In particular, by generalizing the configuration of Ref.~\cite{Dimer07} to two cavity field modes and operating in a dispersive regime (amounting to far-off-resonant Raman transitions), we find that it is possible to implement atomic spin systems that are described by the most general LMG model~(\ref{LMGHamiltonian}), and for which the Hamiltonian dynamics may still dominate over losses to the output cavity fields, thus enabling the clear realization of critical phenomena, including both first- and second-order dynamical quantum phase transitions. We find also that the cavity output fields can again be used to provide clear and detailed probes of  properties of the atomic collective-spin system, including entanglement, in the critical regime.

We note that the present work bears some relation to studies of optical bistability and resonance fluorescence in cooperative atomic systems, which can also exhibit first- and second-order nonequilibrium phase transitions (see, for example, \cite{Bonifacio76,Drummond78,Walls78,Drummond80,Carmichael80}). There, however, the dynamics explicitly includes (resonant) coherent driving of the atomic system by an external laser field (i.e., the Hamiltonian describing the system contains a driving term linear in $J_x$ or $J_y$, rather than a direct spin-spin interaction term), and relatively little investigation has been made of the quantum entanglement associated with the critical behavior~\cite{Schneider}.

A more specific outline of our paper is as follows.
In Sec.~\ref{sect:general_model} we describe the microscopic model of atoms and light fields that realizes our effective spin system. In Sec.~\ref{sect:spin_models} we present some background to the LMG collective spin model and show how to engineer it using the general setup presented in Sec.~\ref{sect:general_model}. We conclude Sec.~\ref{sect:spin_models} with a brief overview of the methods of analysis to be used later in the paper. In Sec.~\ref{sect:implementation} we describe a more specific, potential physical implementation of the system we have proposed, based on alkali metal atoms confined within a high-finesse ring cavity. In Sec.~\ref{sect:ferro_model} we focus on the $\gamma=0$ LMG model and focus on the second-order transition; we first present a linearized analysis of the system in the thermodynamic limit using the Holstein-Primakoff representation of spin operators. Using the input-output theory of quantum optics we relate the internal spin properties to the measurable cavity output field and determine the probe transmission spectrum as an example. The second part of Sec.~\ref{sect:ferro_model} is concerned with the presence and behavior of atom-atom (or spin-spin) entanglement in the system, particularly across the quantum phase transition (QPT). We present results for both the steady-state entanglement and the entanglement dynamics, using either exact numerical solutions for finite system size or analytical solutions in the thermodynamic limit. In Sec.~\ref{sect:af_model} we essentially repeat the analysis of the previous section, but focus on a parameter regime where a first-order phase transition occurs in the $\gamma=0$ LMG model as the effective magnetic field parameter, $h$, is varied.
Finally, in Sec.~\ref{sect:conclusion} we conclude and briefly discuss possible extensions of the current work.


\section{Theoretical Model} \label{sect:general_model}

\begin{figure}[h!t]
\centerline{\includegraphics[width=8.6cm]{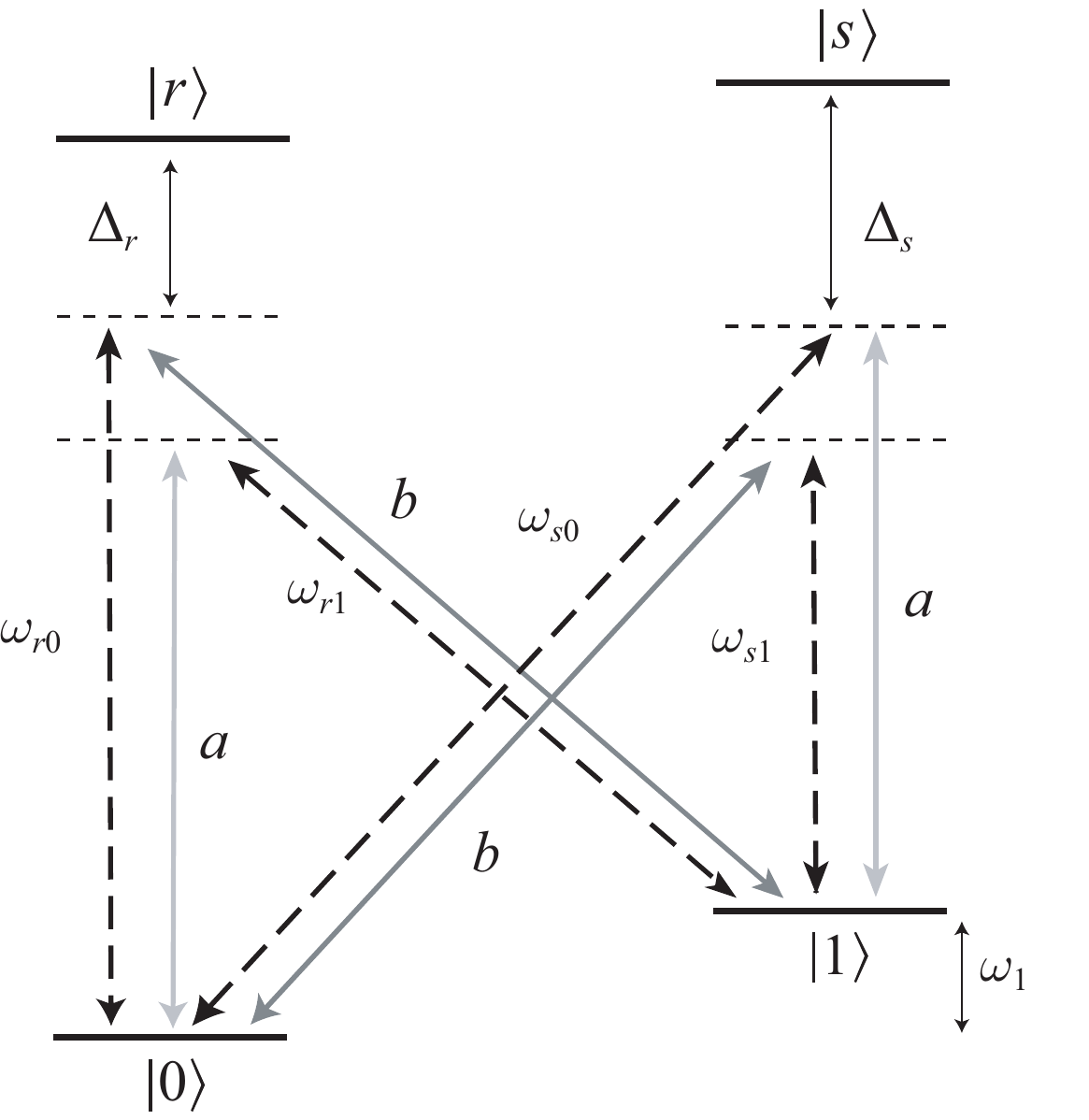}}
\caption{Atomic level and excitation scheme for the general model.} \label{fig:atomic_level_scheme}
\end{figure}

We consider a collection of $N$ atoms coupled via electric dipole transitions to (at most) four laser fields and to a pair of orthogonally polarized optical cavity modes. The atomic level and excitation scheme is shown in Fig.~\ref{fig:atomic_level_scheme}. In particular, the atoms are assumed to possess two stable electronic ground states, labelled $\ket{0}$ and $\ket{1}$, at energies ($\hbar=1$) $\omega_0=0$ and $\omega_1$, respectively. The laser and cavity fields combine to drive Raman transitions between $\ket{0}$ and $\ket{1}$, via the excited atomic states $\ket{r}$ and $\ket{s}$ (energies $\omega_r$ and $\omega_s$, respectively). Specifically, the laser fields, at frequencies $\omega_{r0}$, $\omega_{s0}$, $\omega_{r1}$, and $\omega_{s1}$, couple to the dipole transitions $\ket{0} \leftrightarrow \ket{r}$, $\ket{0} \leftrightarrow \ket{s}$, $\ket{1} \leftrightarrow \ket{r}$, $\ket{1} \leftrightarrow \ket{s}$ with Rabi frequencies $\Omega_{r0}$, $\Omega_{s0}$, $\Omega_{r1}$, and $\Omega_{s1}$, respectively. Cavity field $a$, at frequency $\omega_a$, couples to the transitions $\ket{0} \leftrightarrow \ket{r}$ and $\ket{1} \leftrightarrow \ket{s}$ with coupling strengths $g_{r0}$ and $g_{s1}$, respectively, while cavity field $b$, at frequency $\omega_b$, couples to the transitions $\ket{0} \leftrightarrow \ket{s}$ and $\ket{1} \leftrightarrow \ket{r}$ with coupling strengths $g_{s0}$ and $g_{r1}$, respectively. All of the fields will be assumed to be far-off resonance with the electric dipole transitions to which they couple, meaning that the atomic states $\ket{r}$ and $\ket{s}$ are only virtually excited and can be eliminated from the dynamics.
Finally, at the location of the atoms, the cavity and laser fields are taken to be travelling waves copropagating in the $x$ direction, with sufficiently broad beam waists so as to ensure a homogeneous atom-field coupling.

\subsection{Adiabatic elimination of atomic excited states}

To facilitate adiabatic elimination of the atomic excited states we move to a rotating frame according to the unitary transformation $U(t) = e^{-iH_0 t}$, with
\begin{eqnarray}
H_0 & = & (\omega_{s0} - \omega_1') a^\dagger a + (\omega_{r0} - \omega_1') b^\dagger b \nonumber
\\
&+& \sum_{j=1}^N \left( \omega_{s0} \ketbra{s_j}{s_j} + \omega_{r0} \ketbra{r_j}{r_j}
+ \omega_1'\ketbra{1_j}{1_j} \right) ,
\end{eqnarray}
where $\omega_1'$ is a frequency close (or possibly equal) to $\omega_1$. Next, as mentioned above, we assume large detunings of the light fields from the atomic excited states, i.e., we assume that $\Delta_r=\omega_r-\omega_{r0}$ and $\Delta_s=\omega_s-\omega_{s0}$ are much larger in magnitude than any other rates characterizing the system. This allows the atomic excited states to be adiabatically eliminated and also enables us to neglect the effects of atomic spontaneous emission.

Additionally, as depicted in Fig.~\ref{fig:atomic_level_scheme}, we assume that only four distinct Raman transitions are of significance (i.e., resonant or roughly resonant); in particular, in our model we retain only those Raman processes that cause a change in the electronic state of the atoms ($\ket{0}\rightarrow\ket{1}$ or $\ket{1}\rightarrow\ket{0}$) and also involve transfer of a photon from a laser field into a cavity mode or vice versa. All other possible Raman processes are assumed to be far-off resonant and therefore negligible. Quantitatively, this requires, for example, that $|\omega_{r0}-\omega_a|$ and $|\omega_{r0}-(\omega_{r1}+\omega_1)|$ are sufficiently large, with, in particular, $|\omega_{r0}-\omega_a|,|\omega_{r0}-(\omega_{r1}+\omega_1)|\gg |\omega_a-(\omega_{r1}+\omega_1)|,|\omega_{r0}-(\omega_b+\omega_1)|$.

Retaining only the four dominant Raman processes simplifies the model considerably, and with a choice of laser frequencies such that $\omega_{s0}-\omega_{r1}=\omega_{r0}-\omega_{s1}=2\omega_1'$ we are able to remove all explicit time dependence from the Hamiltonian describing our system. Employing the collective spin operators,
\begin{eqnarray}
J_z &=& \frac{1}{2} \sum_{j=1}^N \left(\ketbra{1_j}{1_j} - \ketbra{0_j}{0_j}\right),
\\
J_+ &=& \sum_{j=1}^N \ketbra{1_j}{0_j} , ~~~
J_- = \sum_{j=1}^N \ketbra{0_j}{1_j} ,
\end{eqnarray}
and omitting constant energy terms, our effective Hamiltonian for the collective atomic system and cavity modes can be written in the form
\begin{eqnarray}
H_{\rm g} &= &\omega_0 J_z +\delta_a a^\dagger a + \delta_b b^\dagger b + 2\delta_a^- J_z a^\dagger a + 2\delta_b^- J_z b^\dagger b \nonumber \\
&&  +  \frac{\lambda_a}{\sqrt{N}} (X_a a+X_a^\dagger a^\dagger) + \frac{\lambda_b}{\sqrt{N}} (X_b b+X_b^\dagger b^\dagger), \label{eq:Heff}
\end{eqnarray}
where $X_i = \alpha_i J_+ + \beta_i J_-$ and the effective parameters are given in terms of the microscopic parameters by
\begin{subequations}
\begin{eqnarray}
\omega_0 & = & \frac{1}{4} \left( \frac{|\Omega_{r1}|^2}{\Delta_r} + \frac{|\Omega_{s1}|^2}{\Delta_s} - \frac{|\Omega_{r0}|^2}{\Delta_r} -\frac{|\Omega_{s0}|^2}{\Delta_s} \right)
\nonumber
\\
&& ~~~~ +\, \omega_1 - \omega_1' , \label{eq:def(a)}
\\
\delta_a &=& \omega_a - \omega_{s0} + \omega_1'+N\delta_a^+ ,
\\
\delta_b &=& \omega_b - \omega_{r0} + \omega_1'+N\delta_b^+ ,
\\
\delta_a^\pm & = &\frac{1}{2} \left( \frac{|g_{s1}|^2}{\Delta_s} \pm \frac{|g_{r0}|^2}{\Delta_r} \right), \label{eq:def(b)}
\\
\delta_b^\pm & = &\frac{1}{2} \left( \frac{|g_{r1}|^2}{\Delta_r} \pm \frac{|g_{s0}|^2}{\Delta_s} \right), \label{eq:def(c)}
\\
\lambda_a \alpha_a & = & \frac{\sqrt{N}\Omega_{r1}^*g_{r0}}{2\Delta_r}, ~~~ \lambda_a\beta_a = \frac{\sqrt{N}\Omega_{s0}^*g_{s1}}{2\Delta_s}, \label{eq:def(d)}
\\
\lambda_b \alpha_b & = & \frac{\sqrt{N}\Omega_{s1}^*g_{s0}}{2\Delta_s}, ~~~ \lambda_b\beta_b = \frac{\sqrt{N}\Omega_{r0}^*g_{r1}}{2\Delta_r}. \label{eq:def(e)}
\end{eqnarray}
\end{subequations}
Note that the (dimensionless) factors $\{\alpha_{a,b},\beta_{a,b}\}\in [-1,1]$ have been introduced for convenience.
Note also that for a characteristic level scheme as shown in Fig.~\ref{fig:atomic_level_scheme}, one might typically expect that $g_{s1} = g_{r0}$ and $g_{r1} = g_{s0}$, so assuming $\Delta_s \approx \Delta_r$ we would therefore also expect that $|\delta_{a,b}^-|\ll |\delta_{a,b}^+|$.

In summary, the master equation for the reduced density operator, $\rho_{\rm g}$ (i.e., with the atomic excited states eliminated and spontaneous emission neglected), is given by
\begin{eqnarray}
\dot{\rho}_{\rm g} = -i[H_{\rm g},\rho_{\rm g}] + \kappa_a D[a]\rho_{\rm g} + \kappa_b D[b]\rho_{\rm g}, \label{eq:master_eqn_atom_cavity}
\end{eqnarray}
where $D[A]\rho = 2 A \rho A^\dagger - A^\dagger A \rho - \rho A^\dagger A$ and $\kappa_i$ is the
cavity field decay rate.

\subsection{Adiabatic elimination of the cavity modes}

We now consider the limit $\sqrt{\kappa_i^2+\delta_i^2}\gg \lambda_a,\lambda_b,\omega_0$. In this limit, the cavity modes are only ever weakly or virtually excited and may also be adiabatically eliminated from the dynamics. Following the standard adiabatic elimination procedure~\cite{QuantumNoise}, we derive the following master equation for the reduced density operator, $\rho$, of the collective atomic system alone:
\begin{eqnarray}
\dot{\rho} = -i[H,\rho] + \Gamma_a D[X_a^\dagger]\rho + \Gamma_b D[X_b^\dagger]\rho,  \label{eq:spin_master_eqn_general}
\end{eqnarray}
with
\begin{equation}
H = \omega_0 J_z - \frac{\Lambda_a}{N} X_aX_a^\dagger - \frac{\Lambda_b}{N} X_bX_b^\dagger,  \label{eq:spin_Hamiltonian_general}
\end{equation}
where the effective spin-spin interaction strengths and collective atomic dissipative rates are
($i\in\{a,b\}$)
\begin{subequations}
\begin{eqnarray}
\Lambda_i &=& \frac{\lambda_i^2\delta_i}{\kappa_i^2+\delta_i^2} \label{eq:Lambda_i} ,
\\
\Gamma_i &=& \frac{\lambda_i^2\kappa_i}{\kappa_i^2+\delta_i^2}  \label{eq:Gamma_i} .
\end{eqnarray}
\end{subequations}
Note that both dispersive nonlinear terms [terms proportional to $\delta_a^-$ and $\delta_b^-$ in Eq.~(\ref{eq:Heff})] do not contribute in the adiabatic approximation since in this limit we assume a vacuum state for both cavity modes.

\subsection{Cavity output fields and measurement} \label{sect:measurement}

Taking a brief step backwards now to the atom-cavity Hamiltonian~(\ref{eq:Heff}), and using the input-output theory of open quantum optical systems~\cite{QuantumNoise,CollettGardiner84_85}, we can derive quantum Langevin equations for the cavity mode operators; in particular, for the mode $b$, we have (neglecting the term proportional to $\delta_b^-$)
\begin{eqnarray}
\dot{b} = -(\kappa_b+i\delta_b) b -i\lambda_b \frac{X_b^\dagger}{\sqrt{N}} + \sqrt{2\kappa_b}\, b_{\rm in}(t),
\label{eq:QLE}
\end{eqnarray}
where $b_{\rm in}(t)$ describes the quantum noise input to the cavity mode (see Fig.~\ref{fig:implementation}) and satisfies the commutation relation $[b_{\rm in}(t),b_{\rm in}^\dag (t')]=\delta(t-t')$. Equation~(\ref{eq:QLE}) illustrates the linear relationship between the cavity operator and atomic operator $X_b^\dag$. The adiabatic limit of the preceding subsection amounts, in the present context, to the assumption that $X_b(t)$ varies on a much slower time scale than $b(t)$ [and $b_{in}(t)$], so that we can write
\begin{eqnarray}
b(t) \simeq -i\frac{\lambda_b}{\kappa_b+i\delta_b} \frac{X_b^\dagger(t)}{\sqrt{N}}
+ \frac{\sqrt{2\kappa_b}}{\kappa_b+i\delta_b}\, b_{\rm in}(t) .
\end{eqnarray}
The cavity output field is given by $b_{\rm out}(t) = \sqrt{2\kappa_b}\, b(t) - b_{\rm in}(t)$, so we in turn obtain a direct relationship between the dynamics of the (internal) collective atomic spin and the (external) cavity output field. Hence, spin-spin correlations of the form $\av{X_bX_b}/N$ and $\av{X_b^\dag X_b}/N$ could be deduced from correlations of the cavity output field, which may be measured, for example, by performing broadband homodyne detection on the emitted light~\cite{WisemanQuadrature}.


\section{Collective (LMG) Spin Models} \label{sect:spin_models}

The LMG model, originally introduced in nuclear physics to model collective motion in nuclei~\cite{originalLMG123}, describes $N$ interacting fermions distributed on two $N$-fold degenerate levels (denoted by $\pm$) separated by an energy $\delta$. Denoting the fermion annihilation operator by $c_{j,\sigma}$, where $j\in\{1,\ldots ,N\}$ and $\sigma\in\{+,-\}$, the Hamiltonian for this system may be written as
\begin{eqnarray}
H' &=& \frac{\delta}{2} \sum_{j,\sigma}\sigma c^\dagger_{j,\sigma} c_{j,\sigma} + \frac{V}{2} \sum_{j,j',\sigma} c^\dagger_{j,\sigma} c^\dagger_{j',\sigma} c_{j,-\sigma} c_{j',-\sigma} \nonumber
\\
&& + \frac{W}{2} \sum_{j,j',\sigma} c^\dagger_{j,\sigma} c^\dagger_{j',-\sigma} c_{j,-\sigma} c_{j',\sigma}.
\end{eqnarray}
Introducing the collective spin operators, $J_z = \frac{1}{2}\sum_{j,\sigma}\sigma c^\dagger_{j,\sigma} c_{j,\sigma}$ and $J_\pm = \sum_j c^\dagger_{j,\pm} c_{j,\mp}$ allows us to reexpress the Hamiltonian as
\begin{eqnarray}
H' &=& \delta J_z  + \frac{V}{2} (J_+^2 + J_-^2) + \frac{W}{2} (J_+ J_- + J_- J_+).
\end{eqnarray}
This Hamiltonian commutes with ${\bf J}^2$, thus conserving the total angular momentum, and with $e^{i\pi J_z}$, corresponding to a parity (spin-flip) symmetry~\cite{EntLMGSecondOrder}. It is straightforward to rewrite this Hamiltonian in terms of $J_x$ and $J_y$, defined via $J_\pm = J_x \pm iJ_y$, giving the generalized LMG model,
\begin{equation}
H_\textrm{LMG} =  -2 h J_z  - \frac{2\lambda}{N} (J_x^2 +\gamma J_y^2),
\end{equation}
where $\lambda =-(V+W)N/2$, $\gamma = (W-V)/(V+W)$ (we will only consider $\gamma \in [-1,1 ]$), and $h = -\delta/2$.

This model is well known for its second-order symmetry breaking phase transition in the ferromagnetic regime ($\lambda>0$)~\cite{EntLMGCUT}. For small interaction strength the system is in the normal phase, where the ground state is unique and polarized in the direction of the magnetic field. As the interaction is increased above a critical value, $\lambda_{\rm c}$, the system enters the broken phase, where the ground state becomes doubly degenerate and macroscopically displaced from its original configuration, thus breaking the parity symmetry. For the special case $\gamma=1$ the Hamiltonian also commutes with $J_z$, thus enabling a direct analytic solution. All other cases $\gamma \neq 1$ lie in a separate universality class. In the antiferromagnetic regime, $\lambda<0$, the model exhibits a first-order phase transition as the effective magnetic field $h$ crosses $h_{\rm c} = 0$ (provided $\gamma>0$).

Using the setup described in the previous section we can implement the generalized LMG model for any $\gamma$ by making appropriate choices of $\alpha_a, \beta_a,\alpha_b, \beta_b$ in the Hamiltonian~(\ref{eq:spin_Hamiltonian_general}). We now consider three specific cases of general interest.

\subsection{Conventional $\gamma=-1$ LMG model} \label{sub_sect:spin_model_1}

The $\gamma = -1$ LMG Hamiltonian may be implemented by choosing $\alpha_a = \alpha_b = \alpha$ and $\beta_a=-\beta_b=\beta$ (corresponding to $X_a = \alpha J_+ +\beta J_-$ and $X_b = \alpha J_+ -\beta J_-$), and setting $\Lambda_a = -\Lambda_b$ (note that the signs of $\Lambda_{a,b}$ are determined by the signs of the detunings $\delta_{a,b}$), so that
\begin{eqnarray}
H & = & -2hJ_z - \frac{2\lambda}{N} (J_x^2-J_y^2), \label{eq:convlmg(a)}
\end{eqnarray}
with $h = -\omega_0/2$ and $\lambda = 2 \alpha \beta \Lambda_a$. This instance of the LMG model has been most widely studied for its phase transition properties. For the dissipative terms we assume, for simplicity, that $2\Gamma_a = 2 \Gamma_b \equiv \Gamma$, so that the full master equation reduces to the form
\begin{eqnarray}
\dot{\rho} & = & -i[H,\rho] + \frac{\Gamma_+}{N} D[J_+]\rho + \frac{\Gamma_-}{N} D[J_-]\rho, \label{eq:convlmg(b)}
\end{eqnarray}
where $\Gamma_+ = \Gamma \alpha^2$ and $\Gamma_- = \Gamma \beta^2$. The Hamiltonian dynamics can be expected to play a dominant role if $\delta_{a,b}\gg \kappa_{a,b}$ (which corresponds to $|\Lambda_{a,b}|\gg\Gamma_{a,b}$).

\subsection{Isotropic $\gamma=1$ LMG model} \label{sub_sect:spin_model_2}

The isotropic $\gamma = 1$ LMG Hamiltonian may be obtained, for example, by choosing $\alpha_a = \beta_b = 1$ and $\alpha_b=\beta_a = 0$( corresponding to $X_a = J_+$ and $X_b = J_-$), and setting $\Lambda_a = \Lambda_b \equiv \lambda$, which gives
\begin{eqnarray}
H & = & -2hJ_z - \frac{2\lambda}{N} (J_x^2+J_y^2), \label{eq:isolmg(a)}
\end{eqnarray}
where $h = -\omega_0/2$. The full master equation is
\begin{eqnarray}
\dot{\rho} & = & -i[H,\rho] + \frac{\Gamma_a}{N} D[J_-]\rho + \frac{\Gamma_b}{N} D[J_+]\rho \label{eq:isolmg(b)}.
\end{eqnarray}

\subsection{Simple $\gamma=0$ LMG model} \label{sub_sect:spin_model_3}

The $\gamma = 0$ LMG Hamiltonian, which will be focus of our attention in this paper, may be obtained by choosing $\alpha_a = \beta_a = \alpha$ (corresponding to $X_a = 2\alpha J_x$), and setting $\delta_b=0$ (so that $\Lambda_b=0$). This gives
\begin{eqnarray}
H & = & -2hJ_z - \frac{2\lambda}{N} J_x^2, \label{eq:spin_gamma_0_Hamiltonian}
\end{eqnarray}
where $\lambda = 2\alpha^2\Lambda_a$. While the Raman channels involving the cavity mode $b$ could be omitted completely, here we retain one of them (for reasons to be discussed below), with the choice  $\beta_b=\beta$, and $\alpha_b = 0$, corresponding to $X_b = \beta J_-$.
Hence, the full master equation we consider is
\begin{eqnarray}
\dot{\rho} & = & -i[H,\rho] + \frac{\Gamma_a}{N} D[2J_x]\rho + \frac{\Gamma_b}{N} D[J_+]\rho, \label{eq:master_equation_gamma0_lmg_model}
\end{eqnarray}
where the factor $\beta^2$ has been absorbed into $\Gamma_b$.

If we now consider the case where $|\Lambda_a|\gg\Gamma_a$ and $\Gamma_b\gg\Gamma_a$, then the role played by each cavity mode in relation to the atomic system is quite distinct. Specifically, cavity mode $a$ mediates the collective spin-spin interaction required for the Hamiltonian dynamics (with coupling strength $\Lambda_a \simeq \lambda_a^2/\delta_a$), while cavity mode $b$ effectively mediates the collective atomic decay (with rate $\Gamma_b=\lambda_b^2/\kappa_b$). Importantly, we note that $X_b = J_+$ implies a quite direct relationship between moments of the cavity mode operators $b$ and $b^\dag$ and moments of the collective atomic spin operators $J_\pm$; in particular, measurements of the output light field from cavity mode $b$ will provide, rather directly and transparently, characteristic properties of the collective atomic spin.

In contrast, for the $\gamma = -1$ model the two cavity modes mediate the collective spin-spin interaction on an equal footing, i.e., $|\Lambda_a| = |\Lambda_b|$, while the operators $X_a$ and $X_b$ are linear combinations of $J_+$ and $J_-$, which leads to a somewhat less transparent (i.e, arguably less convenient) relationship between correlations of the cavity output fields and atomic spin-spin correlations. Partially for this reason, we focus in this paper on the $\gamma = 0$ model, with a clear distinction between the effective roles of the two cavity modes and a potentially better suitability for measurements of the collective atomic spin properties.

\subsection{Methods of analysis} \label{sect:methods_of_analysis}

To analyze the atomic-spin master equations presented in the preceding sections, we make use of both numerical and analytical techniques. For finite spin $j=N/2$, the master equations can be solved numerically for quite large $N$~\cite{QOToolbox}, owing to the linear scaling of the Hilbert space dimension, $d$, with the system size, i.e., $d = N+1$. In what follows, we will typically present results of numerical simulations for $N\lesssim 100$.

For very large system sizes, $N \gg 1$, it is possible to linearize the quantum fluctuations around the mean spin state (i.e., around the ``Bloch vector''). First we find this mean spin state by calculating the steady-state solutions of the semiclassical equations of motion for the components of the Bloch vector. After a suitable rotation (determined by the mean state) of the spin coordinate system, we use the Holstein-Primakoff (HP) representation of angular momentum operators~\cite{Holstein40,Ressayre75}, which enables a systematic large-$N$ expansion of the master equation, to which we then apply the limit $N\rightarrow\infty$. While all of the results obtained in the linearized regime are exact analytical results, in many cases the expressions obtained are too lengthy to give any useful information; in these cases we simply plot the relevant quantities.


\section{Potential Experimental Implementation} \label{sect:implementation}

For a possible experimental implementation of our scheme, we consider, as mentioned previously, an ensemble of atoms confined inside a high finesse ring cavity that supports two travelling-wave modes, $a$ and $b$. The required laser fields, which are assumed to be at frequencies that are not supported by the resonator, are injected through one of the resonator mirrors so as to be copropagating with the cavity fields through the ensemble.

If we take ${}^{6}{\rm Li}$ as the atomic species, then the atomic level scheme of Fig.~\ref{fig:atomic_level_scheme} can be implemented directly with the two ground magnetic substates $|F=1/2,m=\pm1/2\rangle$ as $|0\rangle$ and $|1\rangle$, and with a magnetic field applied perpendicular to the cavity axis to provide a frequency splitting $2\omega_B$ between these two states. The modes $a$ and $b$ would be orthogonal, linearly polarized cavity modes, with, in particular, mode $a$ polarized along the direction of the magnetic field. (Note that if the two modes happen to be very different in frequency due, for example, to birefringence in the cavity mirrors, then the magnetic field may not be necessary.)

\begin{figure}[h!]
\centerline{\includegraphics[width=8.6cm]{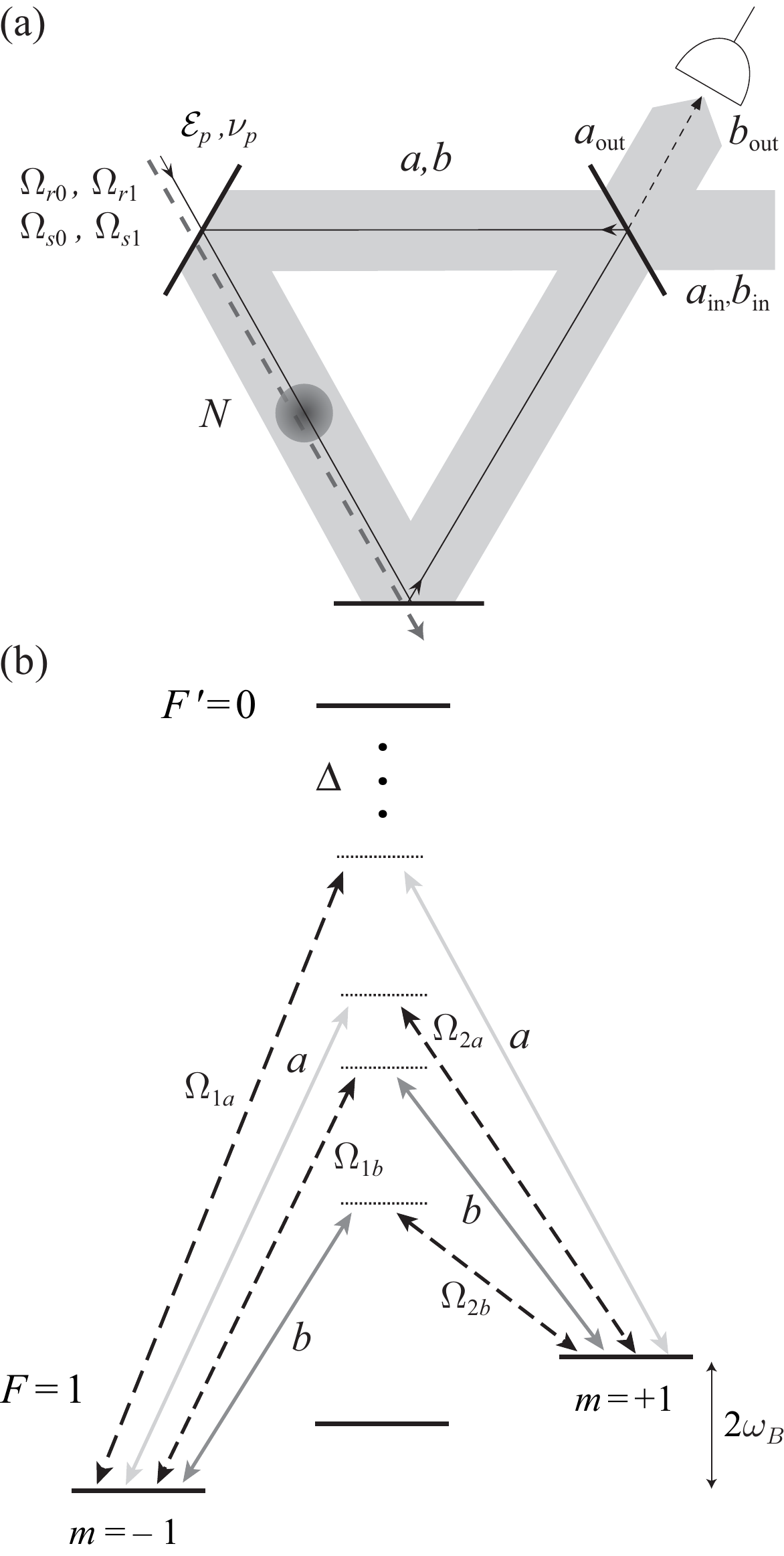}}
\caption{(a) Schematic of potential ring cavity system and setup for measurement of the output transmission spectrum of a weak probe laser field of amplitude ${\cal E}_p$ and frequency $\nu_p$. (b) Possible atomic level scheme as described in the text.} \label{fig:implementation}
\end{figure}

Another possibility, illustrated in Fig.~\ref{fig:implementation}, might be a configuration based on the $F=1\leftrightarrow F'=0$ transition of ${}^{87}{\rm Rb}$, in which the states $|0\rangle$ and $|1\rangle$ are the ground magnetic substates $|F=1,m=\pm1\rangle$, with frequency splitting $2\omega_B$ due to a magnetic field applied along the cavity axis. The modes $a$ and $b$ would be orthogonal, linearly polarized cavity modes, polarized perpendicular to the magnetic field. Note, however, that the modes would need to be sufficiently different in frequency (which could be imposed, for example, by cavity birefringence) in order that the Raman channels involving different modes are distinct.

Alternatively, the modes $a$ and $b$ could be two entirely different (linearly polarized) longitudinal modes of the resonator, one quasiresonant with the $F=1\leftrightarrow F'=0$ transition of the D2 line and the other quasiresonant with the $F=1\leftrightarrow F'=1$ transition of the D1 line.

For specific parameter values, we consider experimental systems such as those realized recently in Ref.~\cite{vonCube06,Klinner06}, where cold atoms are held inside a high-finesse optical ring cavity. In particular, let us assume a single-atom--single-photon dipole coupling strength of $g/(2\pi)\simeq100~\textrm{kHz}$ and a cavity field decay rate of $\kappa_a/(2\pi)\simeq 25~\textrm{kHz}$. For $N\simeq 10^6$ atoms and a characteristic laser-Rabi-frequency-to-detuning ratio of $\Omega/\Delta\simeq 0.005$, we have $\lambda_{a}/(2\pi)\simeq 250~\textrm{kHz}$ ($\alpha_a=1$). If we assume a Raman detuning $\delta_{a}/(2\pi)\simeq 2.5~\textrm{MHz}\gg\lambda_{a}/(2\pi),\kappa_{a}/(2\pi)$, we then have, for example, $\Lambda_{a} \simeq \lambda_{a}^2/\delta_{a} \simeq 2\pi\times 25~\textrm{kHz}$ and $\Gamma_{a} \simeq \Lambda_{a} (\kappa_a/\delta_a) \simeq 2\pi\times 0.25~\textrm{kHz}$. This illustrates that it should be possible to achieve a regime where the (coherent) Hamiltonian dynamics is dominant over the effective dissipation. Note also that, for these parameters, readily achievable ground state magnetic level shifts ($2\omega_B$) of tens of MHz would suffice to ensure distinct Raman channels.

The same parameter regime could obviously be chosen for cavity mode $b$, but if we consider the $\gamma=0$ model as discussed in Sec.~\ref{sub_sect:spin_model_3}, then we might, for example, assume mode $b$ to be more strongly damped (i.e., the two cavity polarizations have different finesses), e.g., $\kappa_{b}/(2\pi)\simeq250~\textrm{kHz}$, and, with smaller Raman transition rate $\lambda_{b}/(2\pi)\simeq 25~\textrm{kHz}$ and detuning $\delta_{b}/(2\pi)\simeq 0$, we would then have $\Gamma_{b} \simeq\lambda_{b}^2/\kappa_{b} \simeq 2\pi\times 2.5~\textrm{kHz}\gg\Gamma_{a}$.
Given these considerations, in the next section, where we examine the second-order transition of the $\gamma=0$ model, we will typically employ the set of normalized parameters $\{h=1,\,\Gamma_a = 0.01,\,\Gamma_b = 0.2\}$, which give a critical coupling strength $\lambda_{\rm c}\simeq 1$.

Finally, we note that the rate for single-atom spontaneous emission (neglected in our model) is estimated by $\Gamma_\textrm{sp}\Omega^2/(4\Delta^2) \lesssim 2\pi \times 0.04~\textrm{kHz} \ll \Lambda_{a},\Gamma_{a,b}$, where an atomic exited state linewidth of $\Gamma_\textrm{sp}/(2\pi) =6~\textrm{MHz}$ has been assumed.


\section{Second-Order Phase Transition} \label{sect:ferro_model}

We focus first on the positive field case ($h>0$) of the $\gamma=0$ LMG model with ferromagnetic interactions ($\lambda>0$), for which a second-order phase transition occurs as the magnitude of the interaction strength is varied~\cite{SecondOrderTransition}. This transition will turn out to be similar to the one recently studied in the dissipative Dicke model with resonant atom-cavity interactions (as considered in Ref.~\cite{Dimer07}). However, it should be noted that in the Dicke model the cavity field plays an intrinsic role in the dynamics and associated critical behavior, unlike in our present model where it has been adiabatically eliminated. Consequently atom-field entanglement is effectively negligible in the present context, while atom-atom entanglement is significant and will be the focus of our study.

In Sec.~\ref{sect:lin_master_eqn} we consider the spin master equation in a linearized regime, appropriate for $N\gg 1$, and determine the transmission spectrum of a weak probe laser. Spin-spin entanglement is studied in Sec.~\ref{sect:entanglement} both in the thermodynamic limit and for finite $N$; specifically the behavior of the steady-state entanglement, as well as entanglement dynamics, is examined in the vicinity of the quantum phase transition.

\subsection{Linearized model} \label{sect:lin_master_eqn}

In this section we study the master equation model~(\ref{eq:master_equation_gamma0_lmg_model}) in the thermodynamic limit by linearizing the quantum fluctuations around the mean-field state. Note that the atom-cavity coupling strengths appearing in the effective coupling constants~(\ref{eq:def(d)}) and~(\ref{eq:def(e)}) scale as $1/\sqrt{V}$, where $V$ is the cavity mode volume. The thermodynamic limit corresponds to $N \rightarrow \infty$ and $V \rightarrow \infty$ with $\varrho = N/V$, the atomic density in the cavity, constant. Since the thermodynamic limit does not alter the effective coupling strengths, which scale as $\sqrt{\varrho}$, we will henceforth refer to the thermodynamic limit as $N\rightarrow \infty$~\cite{Emary03b}.

Firstly, we present the semiclassical analysis which determines the mean-field state relevant for $N\gg 1$. We then expand the angular momentum operators around the semiclassical steady state using the Holstein-Primakoff representation, thus obtaining a linearized version of the master equation, the eigenvalues of which are subsequently analyzed.
Finally, we calculate, for the linearized model, the transmitted amplitude of a weak probe laser through the atom-cavity system as a function of the probe frequency, i.e., the probe transmission spectrum. This physically measurable quantity probes the energy, or eigenvalue, structure of the system and, as we will see, provides clear signatures of the dynamical quantum phase transition.

\subsubsection{Semiclassical equations of motion and steady-state solutions} \label{sect:semiclassical_equations}

The equations of motion for the expectation values of the spin components of the Bloch vector, $\av{J_x}$, $\av{J_y}$, and $\av{J_z}$, are readily derived from the master equation~(\ref{eq:master_equation_gamma0_lmg_model}), but do not form a closed set of equations. However, by factorizing all terms $\av{J_k J_{l}} \rightarrow \av{J_k}\av{J_{l}}$ with $k, l \in \{x,y,z\}$, which corresponds to neglecting quantum fluctuations, we obtain a closed set of equations, which we call the semiclassical equations of motion from hereon.
Introducing the notation $X=\av{J_x}/j$, $Y=\av{J_y}/j$, $Z=\av{J_z}/j$, where $j=N/2$, the semiclassical equations of motion are found to be
\begin{subequations}
\begin{eqnarray}
\dot{X} & = & 2h Y - \Gamma_b Z X,  \label{eq:semicl(a)}\\
\dot{Y} & = & -2h X + 2\lambda Z X - \Gamma_b Z Y, \label{eq:semicl(b)}\\
\dot{Z} & = & - 2\lambda X Y + \Gamma_b(X^2 +Y^2), \label{eq:semicl(c)}
\end{eqnarray}
\end{subequations}
with the constraint $X^2 + Y^2 + Z^2 = 1$ corresponding to conservation of angular momentum.

The steady-state solutions of these equations of motion exhibit a bifurcation at a critical coupling strength
\begin{equation} \label{eq:lambda_c}
\lambda_{\rm c} \equiv h + \frac{\Gamma_b^2}{4h}
\end{equation}
(note $\lambda_{\rm c}>\{h,\Gamma_b\}$ for $\Gamma_b\neq 2h$). For $\lambda <\lambda_{\rm c}$ the stable steady-state solutions are
\begin{eqnarray}
Z_\textrm{ss} =  1 , \quad X_\textrm{ss} =Y_\textrm{ss} = 0, \label{eq:semicl_ss_sols}
\end{eqnarray}
while for $\lambda>\lambda_{\rm c}$ they become
\begin{subequations}
\begin{eqnarray}
Z_\textrm{ss} & = & \frac{2h}{\Lambda}, \label{eq:semicl_ss_sols_(a)} \\
X_\textrm{ss} & = & \pm \sqrt{\frac{\Lambda^2-4h^2}{2\lambda\Lambda}},
\label{eq:semicl_ss_sols_(b)} \\
Y_\textrm{ss} & = & \frac{\Gamma_b}{2h} X_\textrm{ss} Z_\textrm{ss}, \label{eq:semicl_ss_sols_(c)}
\end{eqnarray}
\end{subequations}
where
\begin{equation}
\Lambda = \lambda + \sqrt{\lambda^2-\Gamma_b^2} \, .
\end{equation}

\begin{figure} [h!t]
\centerline{\includegraphics[width=8.6cm]{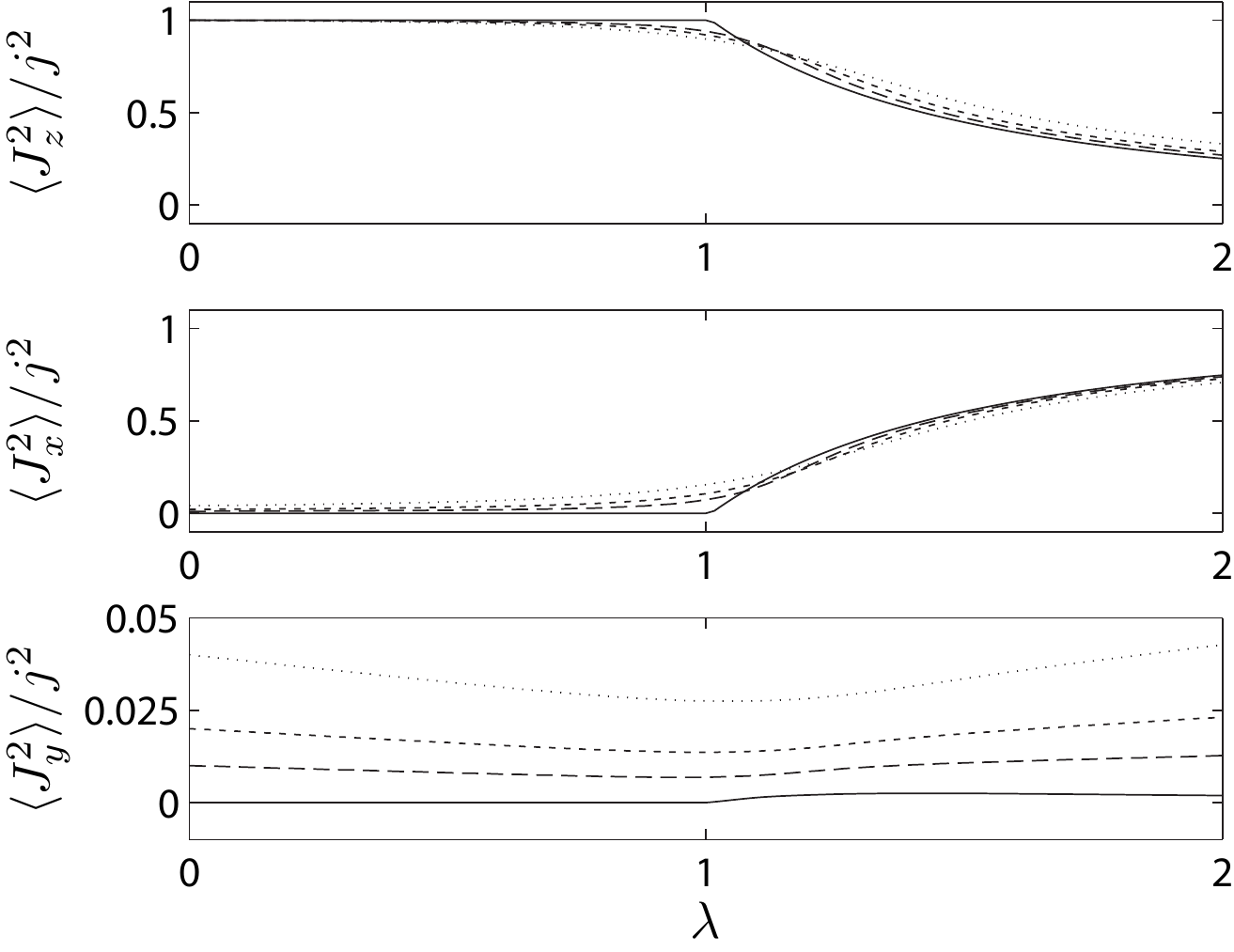}}
\caption{Semiclassical (solid line) and finite-$N$ steady-state second-order moments for $h=1$, $\Gamma_a = 0.01$, $\Gamma_b=0.2$, and $N = 25$ (dotted), $50$ (short dashed line), $100$ (long dashed line).} \label{fig:semicl}
\end{figure}

The bifurcation at $\lambda_{\rm c}$ is illustrated in Fig.~\ref{fig:semicl}, where, to facilitate a comparison between semiclassical and finite-$N$ solutions (computed from numerical solution of the master equation), we plot the second-order moments $\av{J_x^2}$, $\av{J_y^2}$, and $\av{J_z^2}$ (since the finite-$N$ master equation gives $\av{J_x}=\av{J_y}=0$ for all $\lambda$). We note that the two approaches are already in reasonable agreement for $N\simeq50$.

\subsubsection{Holstein-Primakoff representation} \label{sect:hp_rep}

The quantum fluctuations that are neglected in the semiclassical analysis can be included in the limit $N\gg1$ as a first-order correction. This is achieved by using the Holstein-Primakoff (HP) representation of the angular momentum operators~\cite{Holstein40,Ressayre75}, which in the present context takes the form
\begin{subequations}
\begin{eqnarray}
J_z &=& \frac{N}{2} -c^\dagger c , \label{eq:HP_J_z} \\
J_+ &=& \sqrt{N} \sqrt{1-\frac{c^\dagger c}{N}} \, c \, , \label{eq:HP_J_+} \\
J_- &=& \sqrt{N} c^\dagger \sqrt{1-\frac{c^\dagger c}{N}} \label{eq:HP_J_-} \, ,
\end{eqnarray}
\end{subequations}
where $c$ and $c^\dagger$ are bosonic annihilation and creation operators, respectively, satisfying $[c,c^\dagger]=1$.
In particular, if $N\gg 1$ and $\av{J_z}\approx N/2$, i.e., $\av{c^\dag c}\ll N/2$ (so that the Bloch vector points essentially along the $z$ axis), then the HP representation of $J_+$ and $J_-$ can be reduced to $J_+\simeq \sqrt{N}\, c$ and $J_-\simeq \sqrt{N}\, c^\dag$, effectively linearizing the dynamics.

In the normal phase ($\lambda <\lambda_{\rm c}$), this approach can be applied immediately since the steady-state solutions $X_\textrm{ss}=Y_\textrm{ss}=0$. However, in the broken phase ($\lambda>\lambda_{\rm c}$), the steady-state solutions $X_\textrm{ss},Y_\textrm{ss}\neq 0$, i.e., the Bloch vector is rotated away from the $z$ axis, and the HP representation is most conveniently applied with respect to the new orientation of the Bloch vector. We do this by first rewriting the semiclassical steady-state solutions in terms of spherical coordinates $\theta$ and $\phi$ as $Z_\textrm{ss}= \cos{\theta}$, $X_\textrm{ss} = \sin{\theta} \cos{\phi}$, and $Y_\textrm{ss} = \sin{\theta} \sin{\phi}$, and then applying a unitary rotation $R = \exp(i\hat{\bf u}\cdot{\bf J}\theta)$ around an axis $\hat{\bf u} = (-\sin{\phi},\cos{\phi},0)$, so that the transformed operators $J'_{l}=R^\dagger J_l R$ describe quantum fluctuations around the semiclassical steady state. The HP representation~(\ref{eq:HP_J_z})-(\ref{eq:HP_J_-}) and subsequent large-$N$ expansion is then applied to the operators $\{ J_l'\}$.

The master equation obtained in this way may be written, for both phases, in the general form
(omitting constant energy terms in the Hamiltonian)
\begin{eqnarray} \label{eq:linearised_master_equation}
\dot{\rho} & = & -i[H_\textrm{lin},\rho] + \Gamma_{+,k} D[c_k^\dagger]\rho + \Gamma_{-,k} D[c_k]\rho \nonumber \\
&& + \Gamma_{+,k}^s \left[ 2c_k\rho c_k +2 c_k^\dagger \rho c_k^\dagger -\{c_k^2+(c_k^\dagger)^2,\rho \} \right] \nonumber\\
&& -i \Gamma_{-,k}^s \left[ -2c_k\rho c_k +2 c_k^\dagger \rho c_k^\dagger -\{-c_k^2+(c_k^\dagger)^2,\rho \} \right]\nonumber , \\ \label{eq:damping_spin_louvillian}
\end{eqnarray}
with
\begin{eqnarray}
H_\textrm{lin} & = & A_{1,k} c_k^\dagger c_k + A_{2,k} \left[c_k^2 + (c_k^\dagger)^2\right]
\nonumber
\\
&& ~~ + iA_{3,k} \left[(c_k^\dagger)^2-c_k^2 \right],
\label{eq:HP_spin_hamiltonian}
\end{eqnarray}
where $k\in \{<,>\}$ and $c_<$ ($c_>$) denotes the bosonic operator for the normal (broken) phase. The coefficients in the normal phase are given by
\begin{subequations}
\begin{eqnarray}
A_{1,<} &=& 2h - \lambda ,
\\
\quad A_{2,<} &=& -\lambda/2 ,
\\
A_{3,<} &=& 0,
\\
\Gamma_{+,<} &=& \Gamma_a ,
\\
\Gamma_{-,<} &=& \Gamma_a + \Gamma_b ,
\\
\Gamma_{+,<}^s &=& \Gamma_a ,
\\
\Gamma_{-,<}^s  &=& 0,
\end{eqnarray}
\end{subequations}
while in the broken phase they are given by
\begin{subequations}
\begin{eqnarray}
A_{1,>} & = & \frac{1}{2\Lambda} \left( -4h^2-3\Gamma_b^2+4\lambda \Lambda\right) ,
\\
A_{2,>} & = & \frac{1}{4\lambda \Lambda} \left( (\Gamma_b^2 -4 h^2)\sqrt{\lambda^2 - \Gamma_b^2} - 4h \Gamma_b^2\right) ,
\\
A_{3,>} & = & \frac{\Gamma_b}{4\lambda\Lambda}\left(-4h^2 +\Gamma_b^2+4h \sqrt{\lambda^2-\Gamma_b^2}\right) ,
\\
\Gamma_{\pm,>} & = &\frac{\Gamma_a}{2\lambda \Lambda}(4h^2 + \Gamma_b^2) + \frac{\Gamma_b}{4\Lambda^2} (\mp 2h +\Lambda)^2 ,
\\
\Gamma_{+,>}^s & = & \frac{\Gamma_a}{2\lambda^2 \Lambda} \left[ (4h^2-\Gamma_b^2)\sqrt{\lambda^2-\Gamma_b^2} +4h\Gamma_b^2\right] \nonumber \\
&& +\frac{\Gamma_b}{4\lambda \Lambda^2}\sqrt{\lambda^2-\Gamma_b^2}(4h\lambda_{\rm c} -2\lambda\Lambda) , ~~
\\
\Gamma_{-,>}^s& = & \frac{\Gamma_a \Gamma_b}{2\lambda^2 \Lambda} (\Gamma_b^2 -4h^2 + 4h \sqrt{\lambda^2 - \Gamma_b^2}) \nonumber \\
& & + \frac{\Gamma_b^2}{4\lambda\Lambda^2} (\Lambda^2 -4h^2) .
\end{eqnarray}
\end{subequations}
Note that the Hamiltonian~(\ref{eq:HP_spin_hamiltonian}) does not contain any terms linear in $c_k$ and $c_k^\dag$, which is a consequence of the applied rotation, and also means that $\av{c_k}_{\rm ss}=\av{c_k^\dag}_{\rm ss}=0$.

While the coefficients for the broken phase are rather complicated, they do simplify considerably in the limit of very large $\lambda$; in particular, for $\lambda\gg h,\Gamma_{a,b}$ one finds $A_{1,>}\simeq 2\lambda$, $A_{2,>} \simeq 0$, and $A_{3,>} \simeq 0$, while $\Gamma_{\pm,>} \simeq \Gamma_b/4$, $\Gamma_{+,>}^s \simeq -\Gamma_b/4$, and $\Gamma_{-,>}^s \simeq 0$. The master equation then corresponds to that of a simple quantized harmonic oscillator coupled to a somewhat unconventional  (squeezed-type) reservoir~\cite{WallsandMilburn}.

\subsubsection{Eigenvalue analysis} \label{sect:eigenvalues}

\begin{figure}[h!t]
\centerline{\includegraphics[width=8.6cm]{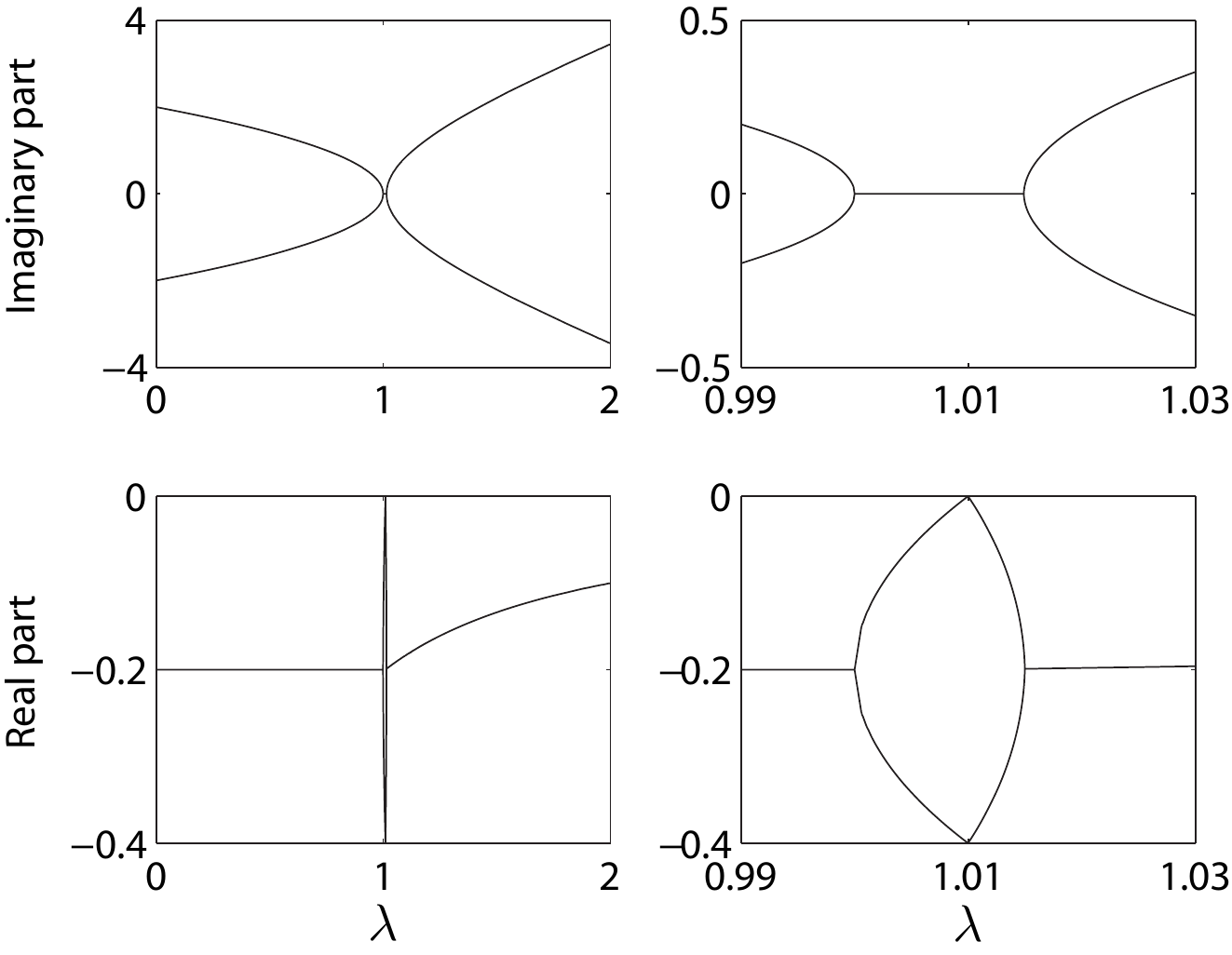}}
\caption{Eigenvalues of the linearized equations of motion, $\mu_\pm$, as given by Eqs.~(\ref{eq:eig_normal_phase}) and~(\ref{eq:eig_broken_phase}), for $h=1$ and $\Gamma_b=0.2$. The right-hand column gives a magnified view of the region around $\lambda_{\rm c}=1.01$.} \label{fig:eig_lin}
\end{figure}

It is interesting to examine the eigenvalues associated with the linear set of equations of motion for the first-order moments $ \av{c_k}, \av{c_k^\dagger}$, which may be expressed as $\dot{\vec{u}} = \mathbf{M} \vec{u}$, where $\vec{u} \equiv (\av{c_k}, \av{c_k^\dagger})^T$ and $\mathbf{M}$ is a $2\times2$ matrix. The real and imaginary parts of these eigenvalues are plotted in Fig.~\ref{fig:eig_lin} for our characteristic set of numerical parameters. We note that except for the region near zero coupling strength the eigenvalues exhibit very similar behavior to that found in the dissipative Dicke model~\cite{Dimer07}.

In the normal phase ($\lambda<\lambda_{\rm c}$) the eigenvalues of $\mathbf{M}$ are given by
\begin{equation} \label{eq:eig_normal_phase}
\mu_\pm = -\Gamma_b\pm 2i \sqrt{h(h-\lambda)} \, ,
\end{equation}
the imaginary parts of which go to zero at the point $\lambda'= h<\lambda_{\rm c}$, with a characteristic scaling of $\sqrt{\lambda'-\lambda}$. For $\lambda'<\lambda<\lambda_{\rm c}$ the eigenvalues are real and distinct, with one going to zero at $\lambda_{\rm c}$ (i.e., critical slowing down) and the other to $-2\Gamma_b$.

In the broken phase ($\lambda>\lambda_{\rm c}$) the eigenvalues of $\mathbf{M}$ are given by
\begin{equation} \label{eq:eig_broken_phase}
\mu_\pm = - \frac{2\Gamma_b h}{\Lambda} \pm \sqrt{2(2 h^2 + \Gamma_b^2 - \lambda\Lambda)} \, .
\end{equation}
In the region $\lambda>\lambda''$, where $\lambda '' =(\Gamma_b^2 + 2 h^2)/\sqrt{4h\lambda_{\rm c}}$, the eigenvalues are complex conjugate pairs with a real part that diminishes for $\lambda \gg \lambda_{\rm c}$ like $-\Gamma_b h/\lambda$.
Provided $\Gamma_b < \sqrt{2}h\sqrt{1+\sqrt{5}}$, then $\lambda''>\lambda_{\rm c}$ and the imaginary parts vanish as $\lambda$ approaches $\lambda''$ from above with the scaling $\sqrt{\lambda - \lambda''}$ (which can be shown using a first-order Taylor series expansion about $\lambda=\lambda''$). The imaginary parts are zero in the interval $\lambda_{\rm c} <\lambda <\lambda''$, while the real parts again approach $0$ and $-2\Gamma_b$, respectively, as $\lambda\rightarrow\lambda_{\rm c}$.

If $\Gamma_b > \sqrt{2}h\sqrt{1+\sqrt{5}}$ then $\lambda ''< \lambda_{\rm c}$, and the eigenvalues are complex conjugate pairs immediately above the critical point.
In this situation, the dissipation is stronger than the Hamiltonian dynamics; this is also an interesting regime, but not one that we will consider in the present paper.

\subsubsection{Probe transmission spectrum} \label{sect:transmission_spectrum}

A standard way to examine the structure and dynamics of an atomic system is to measure the transmission of a (weak) probe laser field through the medium as a function of the probe frequency. This amounts simply to detecting the frequency response of the system to an applied field or ``force''. A schematic diagram illustrating the setup for such a measurement in the present context is shown in Fig.~\ref{fig:implementation} (a).

For our theoretical investigation of the transmission spectrum we retain the two cavity modes in our model and make use of the input-output theory of open quantum systems~\cite{QuantumNoise,CollettGardiner84_85}. In particular, our starting point is the atom-cavity Hamiltonian~(\ref{eq:Heff}) and we again consider the limit $N\gg 1$, so that we can perform a linearization. To do this, we follow our previous working and determine the stable semiclassical steady-state amplitudes of the atom-cavity system from the semiclassical (i.e., factorized) equations of motion for the moments $\{\av{a}, \av{b}, \av{J_x}, \av{J_y}, \av{J_z}\}$. Note that the steady-state cavity mode amplitudes in this approach can be expressed in terms of the atomic amplitudes as (neglecting terms proportional to $\delta_{a,b}^-$ and setting $\delta_b=0$)
\begin{eqnarray}
\frac{\av{a}_\textrm{ss}}{\sqrt{N}} = \frac{-2i\lambda_a}{\kappa_a+i\delta_a}X_\textrm{ss} , ~~
\frac{\av{b}_\textrm{ss}}{\sqrt{N}} = \frac{\lambda_b}{\kappa_b }(Y_\textrm{ss}-iX_\textrm{ss}) .
\end{eqnarray}
Using the HP representation of the atomic spin operators and linearizing about the semiclassical steady states as before leads to the following Hamiltonian for the normal and broken phases,
\begin{eqnarray}
H_{{\rm g},\textrm{lin}} & = & \delta_c c_k^\dagger c_k + \delta_a a_k^\dagger a_k +\delta_b b_k^\dagger b_k \nonumber \\
&& + (A c_k + A^* c_k^\dagger)(a_k+a_k^\dagger) \nonumber \\
&& + (B_1 c_k + B_2 c_k^\dagger)b_k + (B_1^* c_k^\dagger + B_2^* c_k)b_k^\dagger, ~~\label{eq:linearised_atom_cavity_Hamiltonian}
\end{eqnarray}
where $k\in\{<,>\}$, $a_k$ and $b_k$ denote the annihilation operators for the intracavity modes in the normal and broken phases, and the coefficients $\{\delta_c,A,B_1,B_2\}$ are given in Appendix~\ref{sect:appone}.

Employing the quantum Langevin equations of the input-output theory of open quantum systems we can analytically solve for any cavity output correlations and spectra of interest~\cite{Dimer07}. Here, however, we focus simply on the amplitude of a probe laser field transmitted through the system and into the output field, as depicted in Fig.~\ref{fig:implementation} (a). We consider only the case in which a probe laser of frequency $\nu_p$ (in the rotating frame) and amplitude ${\cal E}_p$ drives cavity field mode $b$.

The analytical expression for the amplitude of the transmitted probe, $A_p(\nu)$, is rather complicated, but if we restrict ourselves to a frequency range where $|\nu|\ll \delta_a,\kappa_b$ (also with $\kappa_a \ll \delta_a$), then for $\lambda <\lambda_{\rm c}$ the transmitted probe intensity is well approximated by
\begin{eqnarray}
&& T_p(\nu) = \left| A_p(\nu) \right|^2
\simeq  \left| 1 - \frac{i\Gamma_b}{4\sqrt{h(h-\lambda)}} \right. \nonumber \\
&& \times \left\{\frac{\left( \sqrt{h}+\sqrt{h-\lambda} \right)^2}{\left[ \nu - 2\sqrt{h(h-\lambda)} \right] +i\Gamma_b} \right. \nonumber \\ 
&& \left. \left. -  \frac{\left( \sqrt{h}-\sqrt{h-\lambda} \right)^2}{\left[ \nu + 2\sqrt{h(h-\lambda)} \right] +i\Gamma_b} \right\} \right|^2 , \label{eq:Tp_nu}
\end{eqnarray}
where we have normalized the intensity such that it takes a maximum value of unity for an empty cavity. This form for $T_p(\nu)$ highlights the presence of atomic resonances at the frequencies $\nu = \mathrm{Im} (\mu_\pm)$, superimposed on a broad background corresponding to the bare cavity mode resonance. Note that this is in contrast to the findings in the dissipative Dicke model~\cite{Dimer07} where the probe laser transmission spectrum exhibits strongly coupled atom-cavity resonances.

\begin{figure}[h!t]
\includegraphics[width=8.6cm]{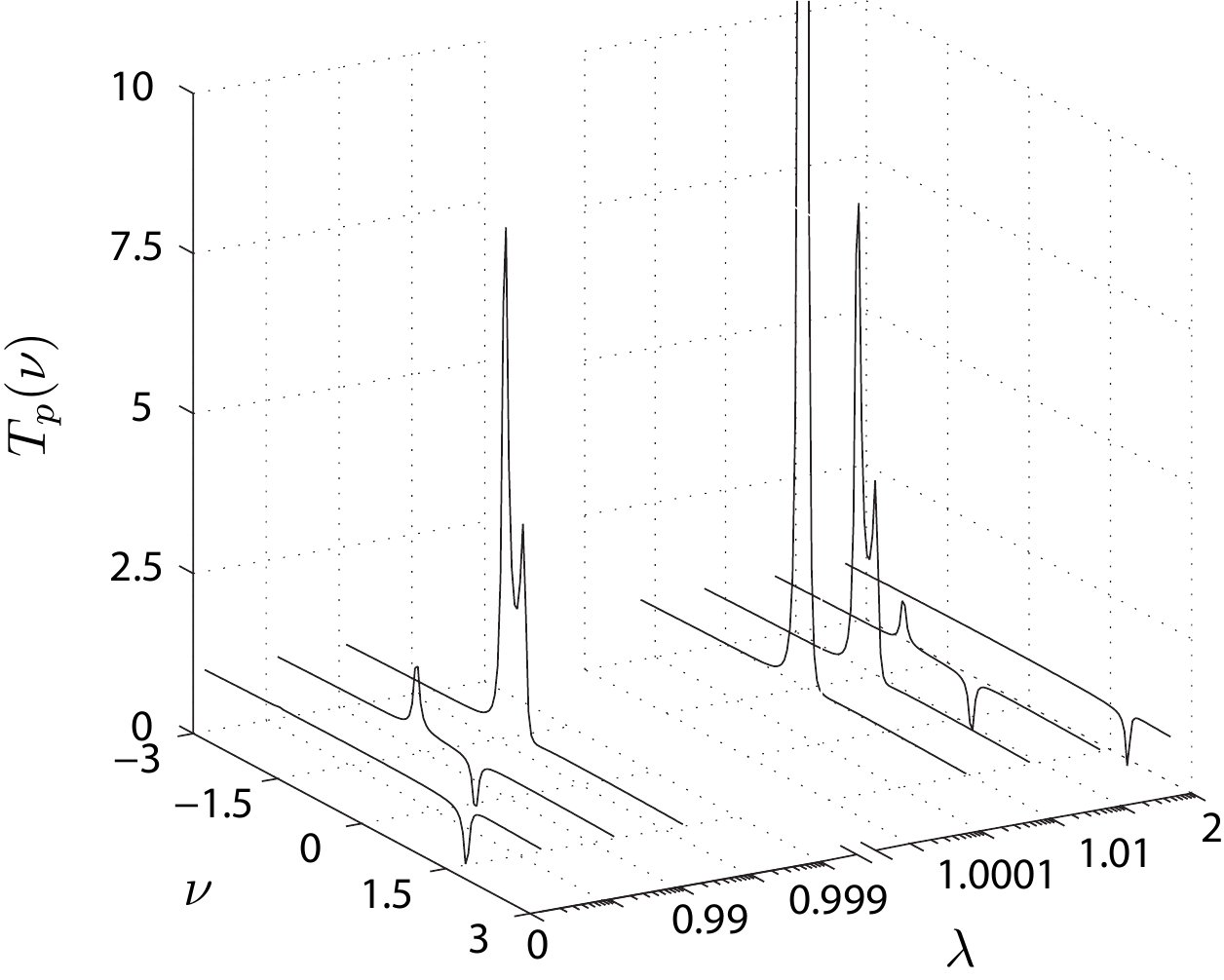}
\caption{Transmission spectrum  in the linearized regime, for $\lambda =0.3, 0.93,0.992, 1.000625 (=\lambda_{\rm c}), 1.005,1.05,1.5 $, with microscopic parameters $\kappa_a =0.3$, $\delta_a = 15$, and $\lambda_b=0.87$, $\kappa_b=15$, giving $\Gamma_b = 0.05$. We set $h=1$ as usual. Note that $\lambda_a$ is chosen to give the indicated $\lambda$ for the given choice of $\kappa_a$ and $\delta_a$, {\it viz}. Eq.~(\ref{eq:Lambda_i}) and recalling that $\lambda = 2 \Lambda_a$, while $\Gamma_a$ varies according to Eq.~(\ref{eq:Gamma_i}), with $\Gamma_a=0.01$ when $\lambda=\lambda_{\rm c}$. } \label{fig:trans_spect}
\end{figure}

In Fig.~\ref{fig:trans_spect} we plot the transmission spectrum [computed from the full theory -- note that the approximate expression~(\ref{eq:Tp_nu}) is in good agreement for the parameters chosen] for a series of values of $\lambda$ on either side or the transition. Note that we have chosen $\Gamma_b = 0.05$ here in order to highlight the main features of the  spectrum more clearly. For $\lambda \ll \lambda_{\rm c}$, we observe, at $\nu\simeq 2h$, a single sharp dip of width $2\Gamma_b$ in the envelope of cavity mode resonance, corresponding to a cavity-mediated collective atomic emission resonance ($\delta_c\simeq 2h$); at this $\lambda$, spin-spin interactions mediated by cavity mode $a$ are small and have little effect on the spectrum.

As the interaction strength $\lambda$ is increased, this dip moves to smaller frequencies and reduces in depth (eventually inverting), while a peak emerges at the corresponding negative frequency. The positions and widths of these features reflect the real and imaginary parts, respectively, of the eigenvalue structure of the system, while their ``intensities'' also relate to the populations of the energy levels. At $\lambda=h$ the two peaks merge into a single peak centered at $\nu=0$, with a height $T_p(0)\simeq (h/\Gamma_b)^2$. Then, as $\lambda\rightarrow\lambda_{\rm c}$, this peak diverges (corresponding to eigenvalue $\mu_-\rightarrow 0$) in a pronounced signature of the phase transition. A similar divergence in the probe laser transmission spectrum is found in the dissipative Dicke model~\cite{Dimer07}.

Just above the critical point, two peaks reappear in the spectrum and move apart with increasing $\lambda$, as shown in Fig.~\ref{fig:trans_spect}. The negative frequency peak diminishes in strength, while the peak at positive frequency inverts to a dip, which narrows and moves to increasingly larger frequencies. In fact, for $\lambda\gg 1$, its position is approximated by $2\lambda$ and its width by $2\Gamma_bh/\lambda$.

\subsection{Entanglement} \label{sect:entanglement}

\subsubsection{Entanglement criteria} \label{sect:entanglement_criteria}

Recently a criterion for bipartite entanglement in collective spin systems was derived~\cite{EntanglementCriteria}, and the connection to spin squeezing established. For the case of symmetric states the criterion is both necessary and sufficient, and reads
\begin{equation}
C_{\varphi} \equiv 1-\frac{4}{N}\av{\Delta J_{\varphi}^2}-\frac{4}{N^2}\av{J_{\varphi}}^2 > 0, \label{eq:finite_N_ent_criteria}
\end{equation}
where $J_\varphi = \sin(\varphi)J_x + \cos(\varphi) J_y $. In this work, we will use the magnitude of $C_\varphi$ as a quantitative measure of the entanglement in the system.
Note that for finite $N$, and also in the linearized analysis, we have $\av{J_{\varphi}}=0$ [since there are no linear driving terms in the effective Hamiltonians~(\ref{eq:spin_gamma_0_Hamiltonian}) and~(\ref{eq:HP_spin_hamiltonian})], and thus $C_{\varphi} = 1-(4/N)\av{J_{\varphi}^2}$. Note also that $C_{\varphi=0} \equiv C_y$, which was shown to be equivalent to the concurrence, $C$~\cite{OrigConcurrece}, in nondissipative LMG models~\cite{EntLMGSecondOrder}.

We also compute the rescaled concurrence, $C_\textrm{R} = (N-1) C$, which is the relevant (nonvanishing) quantity to study for infinitely coordinated collective spin systems in the thermodynamic limit~\cite{Lambert04,EntLMGConcurrReview}. It is possible to show that for the system considered here, $C_\textrm{R}$ may be written as~\cite{MolmerConcurrSymm}
\begin{displaymath}
C_\textrm{R} = \left\{ \begin{array}{lll}
2 \textrm{max} \{0, \mathcal{C}_1 \} & \textrm{if} & E<F  \label{eq:fN_C_R_1} \\
2 \textrm{max} \{0, \mathcal{C}_2 \} & \textrm{if} & E\geq F
\end{array} \right.
\end{displaymath}
where
\begin{eqnarray}
\mathcal{C}_1 & =&  \frac{|\av{J_+^2}|}{N} - \frac{\av{J_x^2} + \av{J_y^2}}{N}+\frac{1}{2} \, , \\
\mathcal{C}_2 & = & \frac{N}{4}-\frac{\av{J_z^2}}{N} \nonumber \\
&& -\frac{\sqrt{[(N(N-2)+4\av{J_z^2}]^2-[4(N-1)\av{J_z}]^2}}{4N} \, , \nonumber \\
\end{eqnarray}
and
\begin{eqnarray}
E & = & \frac{N}{2} - \frac{2\av{J_z^2}}{N} \, , \\
F & = & \frac{\sqrt{[(N(N-2)+4\av{J_z^2}]^2-[4(N-1)\av{J_z}]^2}}{4N}
\nonumber
\\
&& ~~~ +\, \frac{|\av{J_+^2}|}{N} \, .
\end{eqnarray}
As pointed out in Sec.~\ref{sect:measurement}, the spin variances required to compute the entanglement measures described above can in principle be determined from appropriate measurements performed on the cavity output field.

\subsubsection{Steady-state entanglement} \label{sect:steady_state_entanglement}

\begin{figure}[h!t]
\includegraphics[width=8.6cm]{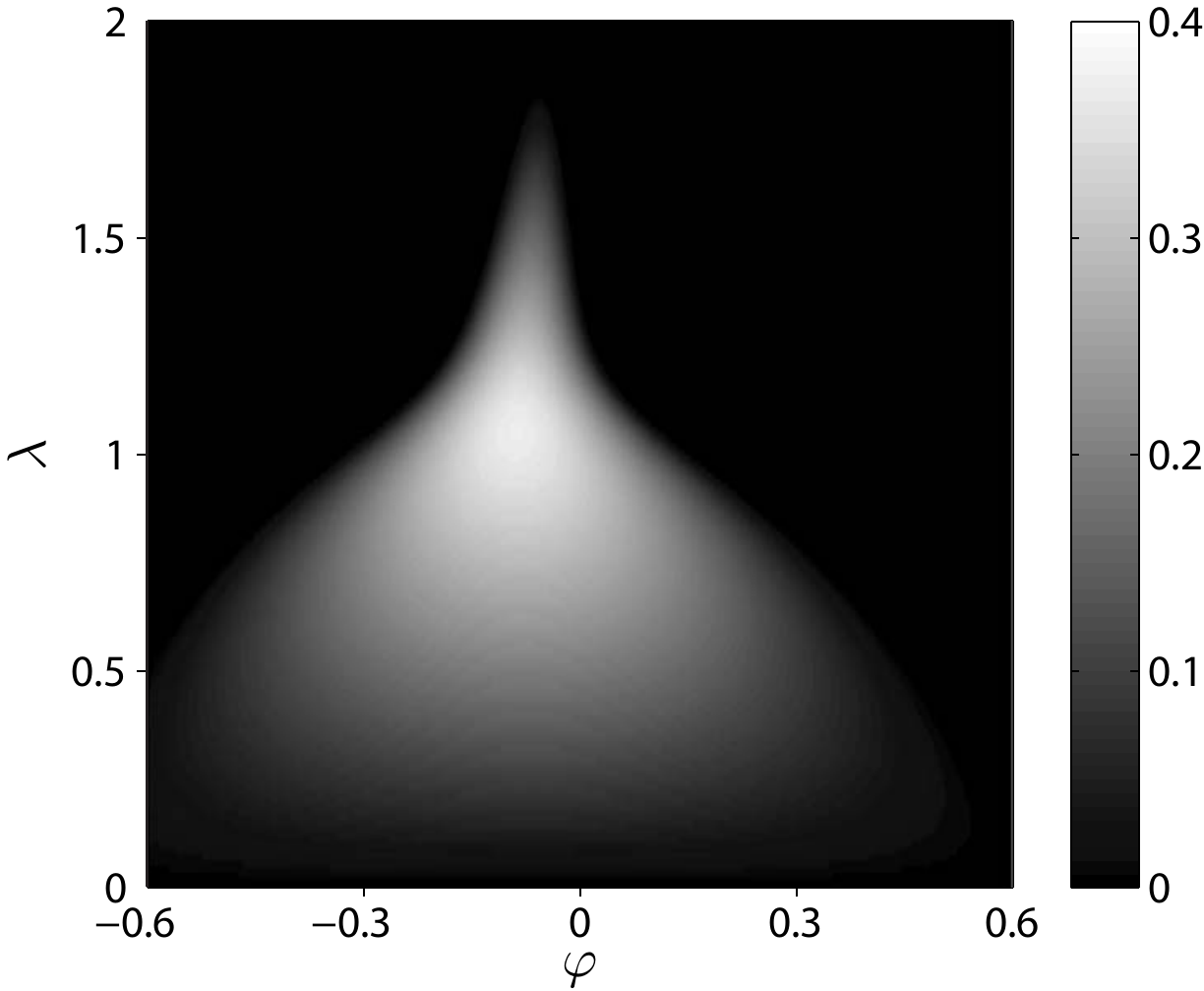}
\caption{Entanglement measure $\textrm{max}\{ 0,C_\varphi\}$ for $N=100$, $h=1$, $\Gamma_a = 0.01$, and $\Gamma_b=0.2$.} \label{fig:fNcssrot}
\end{figure}

For finite $N$ we numerically solve the master equation for the steady-state density matrix and then compute the operator averages required to determine $C_\varphi$ and $C_\textrm{R}$. In Fig.~\ref{fig:fNcssrot} we plot $\textrm{max}\{ 0,C_\varphi\}$ as a function of $\lambda$ and $\varphi$ for $N=100$. We see that below the critical point, $\lambda < \lambda_{\rm c}$, entanglement is present for a broad range of angles $\varphi$. However, as the critical point is approached the range of angles $\varphi$ which gives nonzero entanglement, $C_\varphi>0$, becomes increasingly narrow. Once above the transition, $\lambda>\lambda_{\rm c}$, the region of finite $C_\varphi$ continues to narrow until it eventually disappears altogether.

\begin{figure}[h!t]
\centerline{\includegraphics[width=8.6cm]{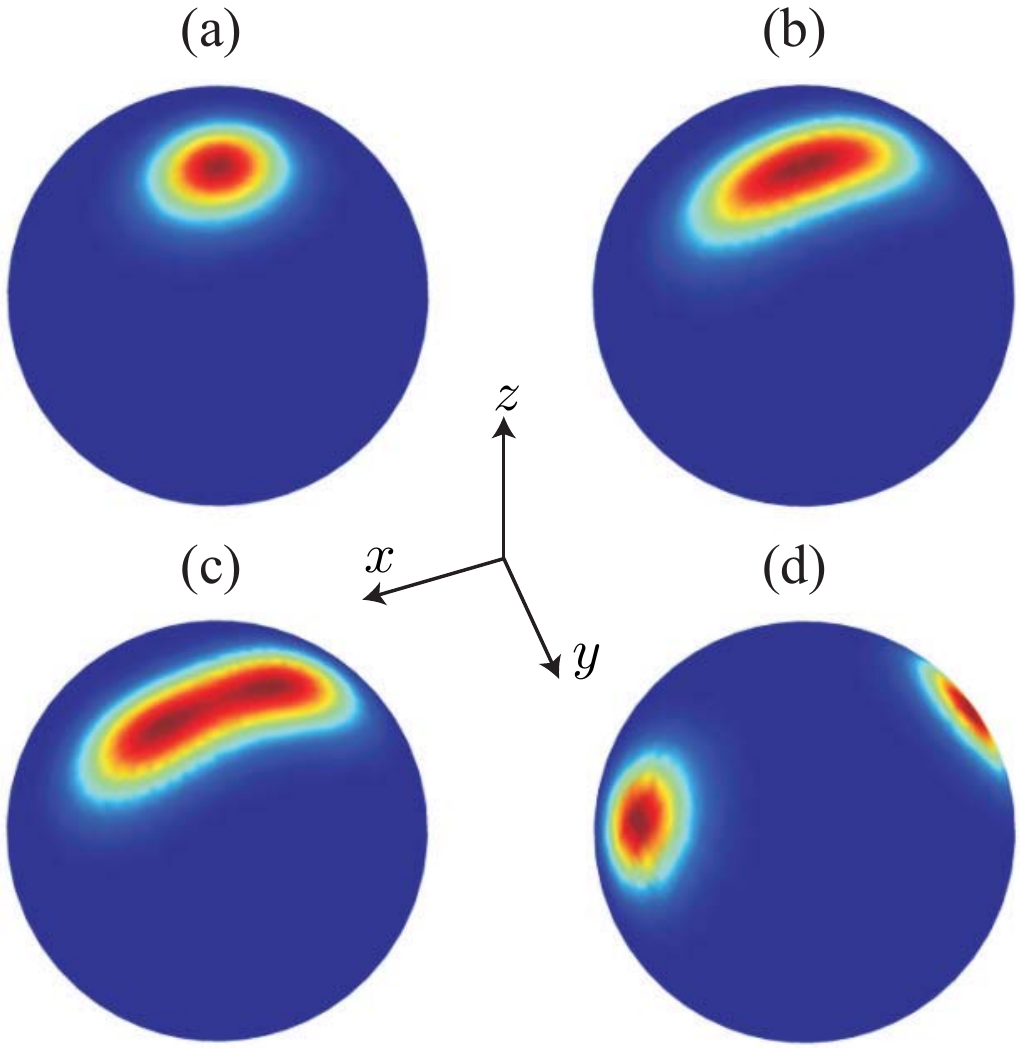}}
\caption{(Color online) Steady-state spin $Q$-function, $Q_\textrm{s}(\eta)$, on the Bloch sphere for (a) $\lambda = 0.5$, (b) $\lambda=1.01$, (c) $\lambda = 1.1$, and (d) $\lambda=2$, with $N=50$, $h=1$, $\Gamma_a = 0.01$, and $\Gamma_b=0.2$. Note that dark blue corresponds to the minimum value of zero of $Q_{\rm s}(\eta)$ while dark red indicates the maximum value of $Q_{\rm s}(\eta)$.} \label{fig:spin_qfunc}
\end{figure}

To help interpret the behavior of $C_\varphi$, we make use of a phase space representation of the atomic state that employs the spin coherent states, which are defined by~\cite{MolmerConcurrSymm}
\begin{eqnarray}
\ket{\eta} = (1+|\eta|^2)^{-j} \sum_{m=-j}^j \sqrt{\binom{N}{j+m}}\eta^{j+m} \ket{j,m}_j \, ,
\end{eqnarray}
where  $\eta = e^{i\phi}\tan{\frac{\theta}{2}}$, with $\theta$ and $\phi$ corresponding to spherical coordinates, and $\ket{j,m}$ are the Dicke states with $m\in [-j,-j+1,\ldots ,j-1,j]$ (for our system, $j=N/2$). Using these states we can define the spin $Q$-function,
\begin{equation}
Q_{\rm s}(\eta ) =\bra{\eta} \rho \ket{\eta}.
\end{equation}
Fig.~\ref{fig:spin_qfunc} displays $Q_{\rm s}(\eta )$ on the surface of the Bloch sphere for $N=50$ and for a series of interaction strengths $\lambda$. Below the critical point, $Q_{\rm s}(\eta )$ is single-peaked and centered around the top of the Bloch sphere ($\theta =0$), with little obvious angular dependence. Correspondingly, the entanglement measure $C_\varphi$ is finite over a rather broad range of $\varphi$, with a maximum close to $\varphi=0$ (i.e., near $C_y$). Note that this slight shift of the optimum away from $\varphi=0$ is a consequence of the dissipation ($\Gamma_b$) in the system.

As $\lambda$ increases towards the critical point, $Q_{\rm s}(\eta )$ becomes increasingly elongated along a direction close to the $x$ axis, until, at the transition, it splits into two peaks located approximately at the two semiclassical steady-state amplitudes~(\ref{eq:semicl_ss_sols_(b)}) and~(\ref{eq:semicl_ss_sols_(c)}). These peaks continue to move apart in phase space as the interaction strength is increased further; eventually both peaks will lie in the equatorial plane corresponding to $\theta=\pi/2$ and $\phi = 0,\pi $. Correspondingly, the range of $\varphi$ over which $C_\varphi$ remains finite becomes increasingly narrow and is focussed around an axis perpendicular to that along which the two peaks lie. This narrowing of the ``width'' of $C_\varphi$ can be explained by noting that, since $\av{J_{\varphi}}=0$, we have $C_{\varphi} = 1-(4/N)\av{[\sin(\varphi)J_x + \cos(\varphi) J_y]^2}$. For increasing interaction strength $\lambda>\lambda_{\rm c}$, $\av{J_x^2}$ becomes of order $j^2=N^2/4$ (see Fig.~\ref{fig:semicl}), and so the optimal choice of $\varphi$ becomes more critical. In fact, one can show for $\lambda>\lambda_{\rm c}$ that the range of $\varphi$ over which $\textrm{max}\{ 0,C_\varphi\}>0$ scales as $1/\sqrt{N}$.

\begin{figure}[h!t]
\includegraphics[width=8.6cm]{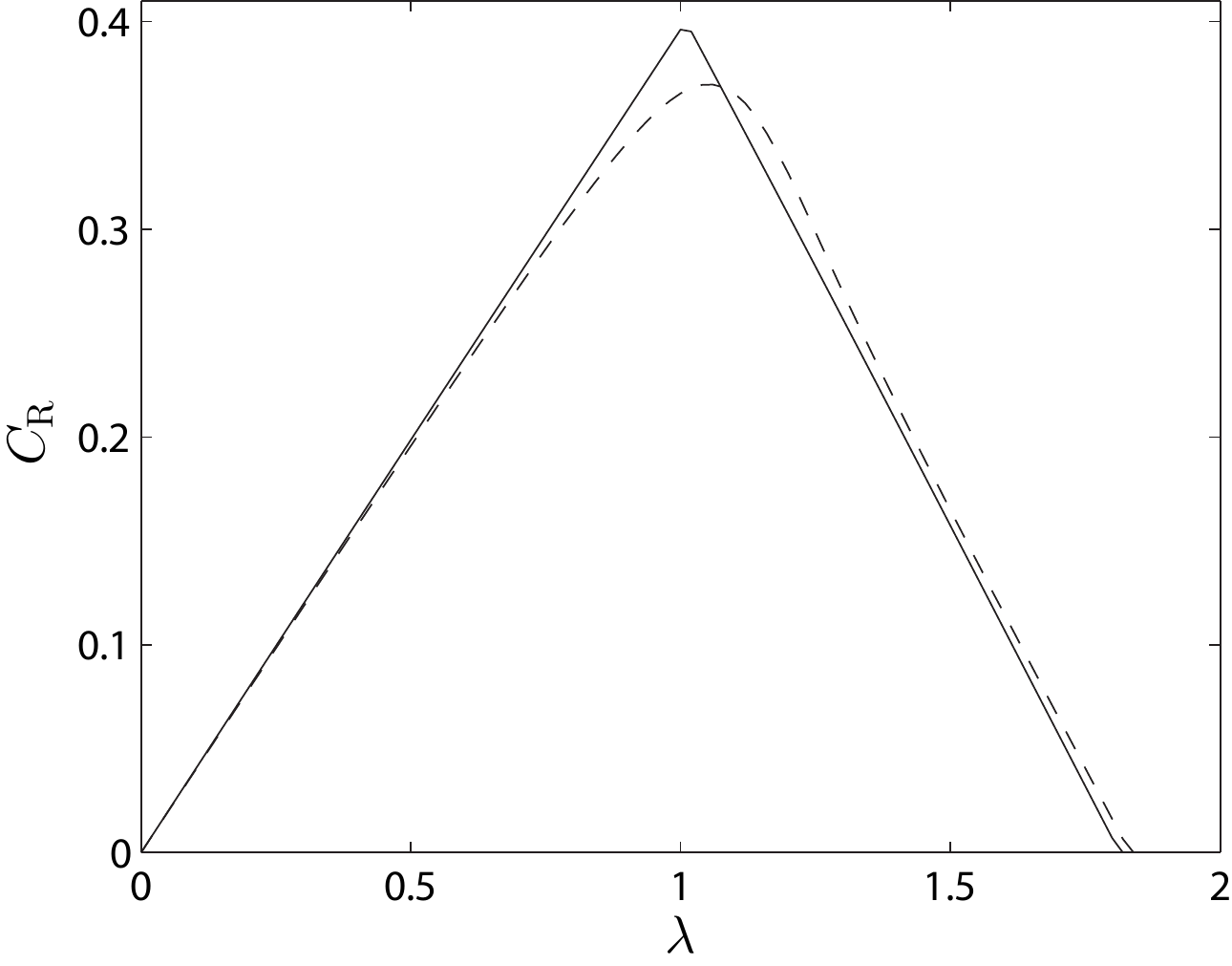}
\caption{Rescaled concurrence $C_\textrm{R}$ for $N=100$ (dashed line) and in the thermodynamic limit (solid line) with $h=1$, $\Gamma_a = 0.01$, and $\Gamma_b=0.2$.}  \label{fig:C_R_second_fN_and_HP_combo}
\end{figure}

Next, we consider the rescaled concurrence, $C_\textrm{R}$, as a function of the interaction strength $\lambda$. In fact, one finds that
\begin{equation} \label{eq:concurr}
C_{\rm R} = \max_\varphi C_\varphi \, ,
\end{equation}
i.e., $C_{\rm R}$ is simply the optimal value of the quantity $C_\varphi$ just considered. In Fig.~\ref{fig:C_R_second_fN_and_HP_combo} we plot $C_\textrm{R}$ versus $\lambda$ and observe that the entanglement reaches a maximum for $\lambda$ close to $\lambda_{\rm c}$ [at finite $N$ the critical point is slightly shifted from $\lambda_{\rm c}$ as given in Eq.~(\ref{eq:lambda_c})]. This peaking of the entanglement at the quantum phase transition has been conjectured and demonstrated theoretically for the equivalent closed (nondissipative) spin models~\cite{EntLMGConcurrReview,EntLMGBlock,EntLMGEntropy}. Our results confirm that this behavior can persist in steady state in our nonequilibrium, open-system version of these models, and can in principle be measured within our proposed setup.

In the linearized treatment ($N\gg1$) of the HP representation, we can write $J_\varphi \approx (\sqrt{N}/2) \chi_\varphi$, where $\chi_\varphi =  i (-c_k e^{i\varphi} + c_k^\dagger e^{-i\varphi})$. Noting that $\av{c_k}=0$, the entanglement measure $C_\varphi$ can be expressed as
\begin{eqnarray}
C_{\varphi}^\textrm{HP} &=& 1-\av{ \chi_{\varphi}^2} \nonumber
\\
&=& \left( e^{2i\varphi}\av{c_k^2} +e^{-2i\varphi}\av{(c_k^\dagger)^2}-2\av{c_k^\dagger c_k} \right) ,
\label{eqn:HP_entcriteria}
\end{eqnarray}
while the rescaled concurrence is $C_{\rm R}$ is given by
\begin{displaymath}
C_\textrm{R}^\textrm{HP} = \left\{ \begin{array}{lll}
2\textrm{max} \{ 0, \mathcal{C}^\textrm{HP}_1 \} & \textrm{if} & E^\textrm{HP}<F^\textrm{HP} \\
2\textrm{max} \{ 0, \mathcal{C}^\textrm{HP}_2 \} & \textrm{if} & E^\textrm{HP}\geq F^\textrm{HP}
\end{array} \right.
\end{displaymath}
where
\begin{eqnarray}
\mathcal{C}^\textrm{HP}_1 & = & |\av{c_k^2}|-\av{c_k^\dagger c_k}, \\
\mathcal{C}^\textrm{HP}_2 & = & \av{c_k^\dagger c_k} - \sqrt{\av{(c_k^\dagger c_k)^2}-\av{c_k^\dagger c_k}} \, ,
\end{eqnarray}
and
\begin{eqnarray}
E^\textrm{HP} &=& 2 \av{c_k^\dagger c_k} ,
\\
F^\textrm{HP} &=& \sqrt{\av{(c_k^\dagger c_k)^2}-\av{c_k^\dagger c_k}}+  |\av{c_k^2}| .
\end{eqnarray}
Using the linearized master equation~(\ref{eq:linearised_master_equation}), we can derive a closed set of equations for the second-order moments $\av{c_k^\dagger c_k}$, $\av{c_k^2}$, and $\av{(c_k^\dagger)^2}$, from which we may determine the steady-state solutions analytically. Note that the fourth-order moment appearing in $F^\textrm{HP}$ can be expressed in terms of second-order moments, since the states we are dealing with in this linearized approximation are necessarily Gaussian.

\begin{figure}[h!t]
\includegraphics[width=8.6cm]{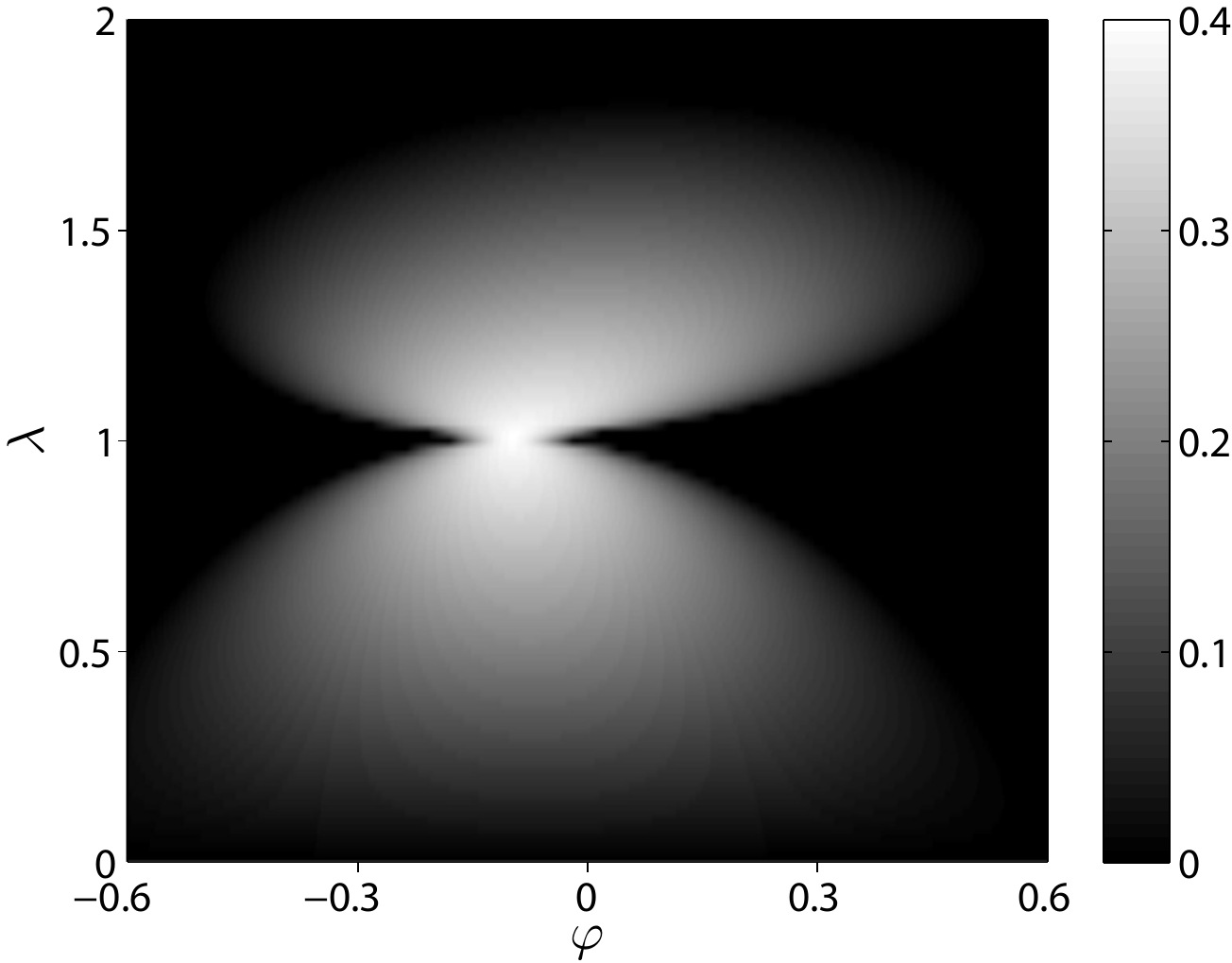}
\caption{Entanglement measure $\textrm{max}\{ 0,C_\varphi\}$ in the thermodynamic limit for $h=1$, $\Gamma_a = 0.01$, and $\Gamma_b=0.2$. Above the critical point, the system is linearized about only one of the two semiclassical steady-state amplitudes; hence the less-sensitive dependence of $\textrm{max}\{ 0,C_\varphi\}$ on $\varphi$ for $\lambda>\lambda_{\rm c}$ as compared with the finite-$N$ results.}  \label{fig:HPcssrot}
\end{figure}

In Fig.~\ref{fig:HPcssrot} we plot $C_\varphi$ as a function of $\varphi$ and $\lambda$, as determined from the linearized HP representation. The behavior below the critical point ($\lambda <\lambda_{\rm c}$) is very similar to the finite-$N$ case. However, the behavior above the critical point ($\lambda>\lambda_{\rm c}$) is very different. Here, the sensitivity of $C_\varphi$ to $\varphi$, for $\lambda >\lambda_{\rm c}$, is much less critical because the linearized model describes only the fluctuations around one of the two semiclassical steady-state amplitudes (i.e., around one of the two lobes appearing in the spin $Q$-function for $\lambda>\lambda_{\rm c}$).

Note that we can obtain plots of $\textrm{max}\{ 0,C_\varphi\}$ similar to Fig.~\ref{fig:fNcssrot} for the region $\lambda>\lambda_{\rm c}$, but determined from the linearized HP model (with a finite value of $N$), by making a rotation back to the original coordinate system and then setting, by hand, $\av{\chi_\varphi}=0$, to mimic an equal, incoherent mixture of the states associated with the two semiclassical amplitudes.

Finally, returning to Fig.~\ref{fig:C_R_second_fN_and_HP_combo}, we have plotted $C_\textrm{R}$ as a function of $\lambda$, computed from the HP model in the thermodynamic limit. Again, $C_\textrm{R}$ corresponds to the value of $C_\varphi$ optimized over $\varphi$, and, since the optimal $\varphi$ corresponds to an axis perpendicular to the (above-transition) splitting of the semiclassical amplitudes, we expect, and indeed find, good agreement with the finite-$N$ results over the full range of $\lambda$.

If we make the simplifying assumption that $\Gamma_a\simeq 0$, then, for $\lambda<\lambda_{\rm c}$ one can show that $E^{\rm HP}<F^{\rm HP}$ and
\begin{eqnarray} \label{eq:HP_CR_Ga_zero}
C_\textrm{R}^\textrm{HP} &\simeq& \frac{\lambda(\sqrt{4h(\lambda_{\rm c}-\lambda)+\lambda^2}-\lambda)}{4h(\lambda_{\rm c}-\lambda)}
\\
&\simeq& \frac{1}{2} - \frac{1}{2}\frac{h(\lambda_{\rm c}-\lambda)}{\lambda_{\rm c}^2} ~~~
\textrm{for} ~~~ \lambda_{\rm c}-\lambda \ll \lambda .
\end{eqnarray}
This shows reasonable agreement with the plot, but reaches a maximum value of 0.5 at the critical point.

\subsubsection{Entanglement dynamics} \label{sect:entanglement_dynamics}

We now consider the dynamics of the entanglement; starting from an initially unentangled state, we examine the time evolution of the state and its entanglement as quantified by $C_\textrm{R}(t)$. The initial state is taken as the $\lambda=0$ ground state, i.e., the state with all atomic spins pointing up (which is a convenient state to prepare in an experiment). As in the previous section, we compute the entanglement both numerically for finite $N$ and analytically for $N\gg 1$ in the linearized approximation.

\begin{figure}[h!t]
\includegraphics[width=8.6cm]{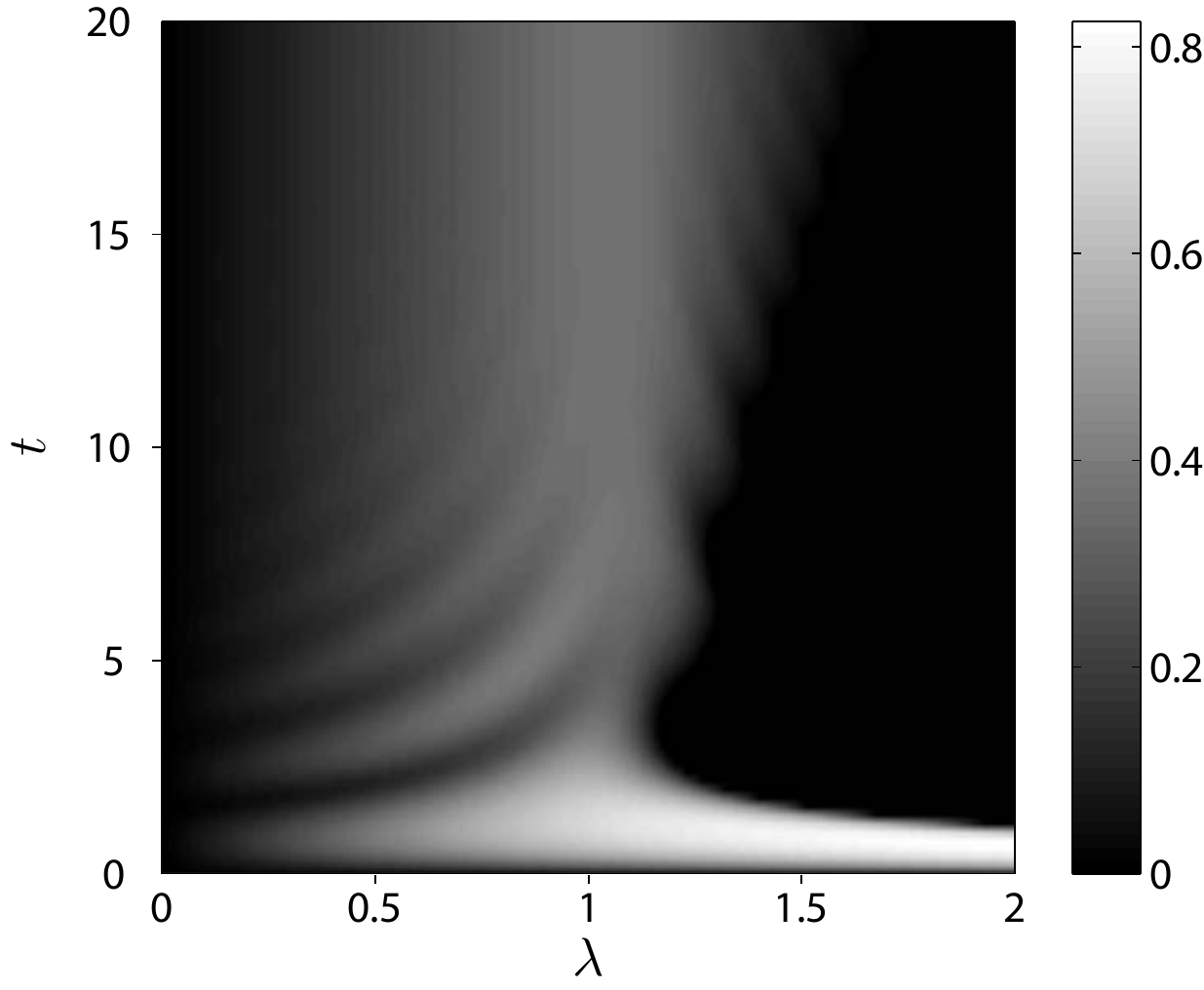}
\caption{Rescaled concurrence $C_\textrm{R}(t)$ as a function of $\lambda$ and $t$ for $N=100$, $h=1$, $\Gamma_a = 0.01$, and $\Gamma_b=0.2$.} \label{fig:finiteNconcurrtimedep}
\end{figure}

In Fig.~\ref{fig:finiteNconcurrtimedep} we plot $C_\textrm{R}(t)$ versus $\lambda$ and time $t$ for $N=100$. At long times we recover the results of the previous section, but at short times the behavior as a function of $\lambda$ is quite different; the entanglement rises to a high value and remains at that value for increasing interaction strength $\lambda$. This behavior can be attributed to the Hamiltonian dynamics, which dominate the dissipation at short times and can create highly entangled states. The potential of such Hamiltonian dynamics for generating such highly entangled states has been proposed previously, for example, in Refs.~\cite{BECMicheli,MolmerGHZ}. Note, however, that the presence of the term $-2hJ_z$ in our system Hamiltonian tends to make the generated states more complicated and less straightforward to interpret. Although the focus of this paper is on the quantum phase transition, it is clear that with a slight modification the scheme also has interesting potential for the controlled generation of specific, highly entangled multiatom states (e.g., Greenberger-Horne-Zeilinger states). In connection with this, an important aspect of our implementation should be highlighted here: because both the effective interaction and dissipation of the spins is controlled by the optical laser fields, we can in principle ``freeze'' the state of the atomic system at any time by simply turning these fields off.

\begin{figure}[h!t]
\includegraphics[width=8.6cm]{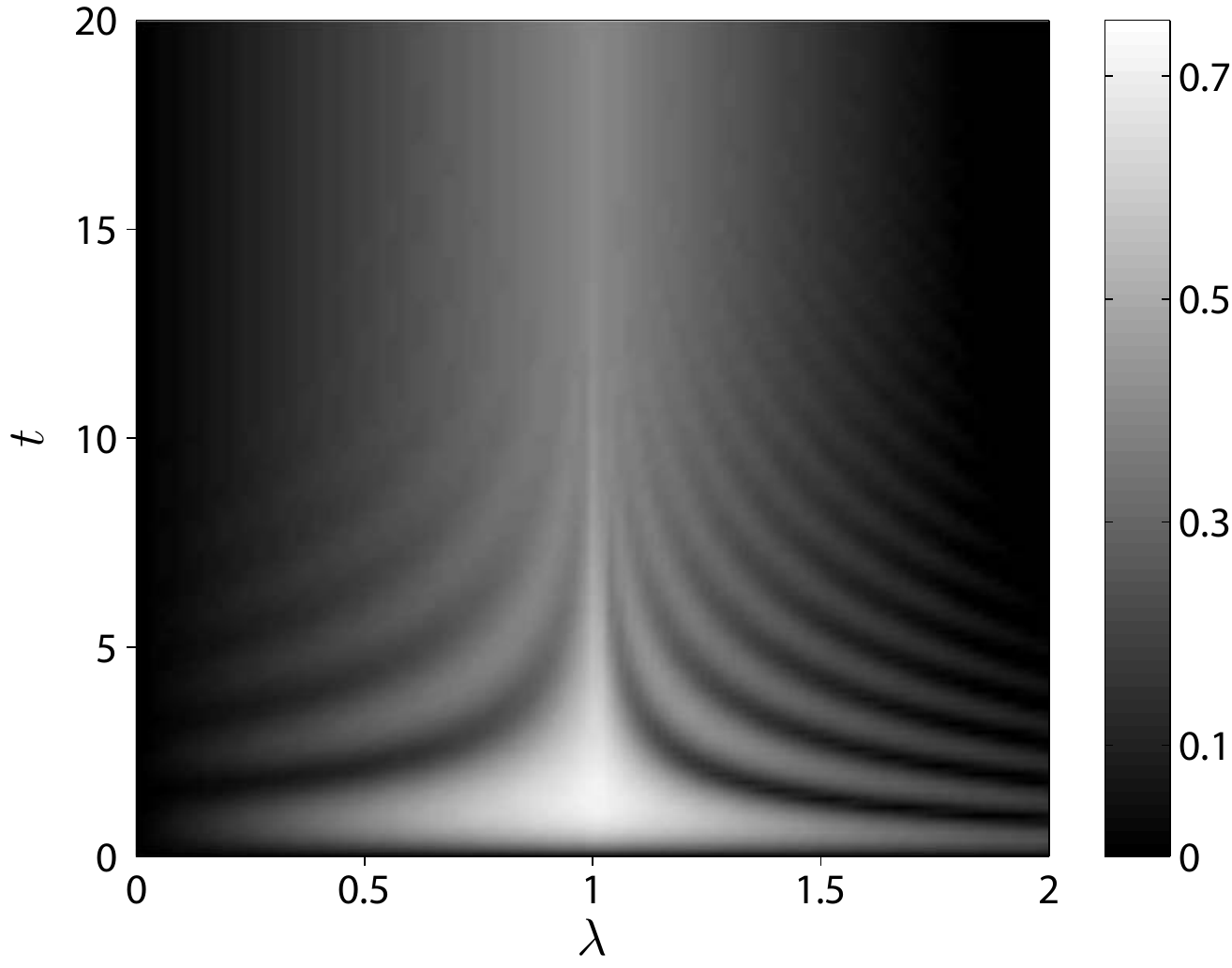}
\caption{Rescaled concurrence $C_\textrm{R}(t)$ as a function of $\lambda$ and $t$ in the thermodynamic limit, with $h=1$, $\Gamma_a = 0.01$, and $\Gamma_b=0.2$. Above the critical point the dynamics is linearized around one of the two possible semiclassical steady-state amplitudes.} \label{fig:HP_C_R_t}
\end{figure}

In the linearized regime, $N\gg1$, we solve the equations of motion for the second-order moments, $\av{c_k^\dagger c_k}$, $\av{c_k^2}$, and $\av{(c_k^\dagger)^2}$, with the initial conditions $\av{c_k^\dagger(0)c_k(0)}=0$, $\av{c_k^\dagger(0)^2}=\av{c_k(0)^2}=0$. The results for $C_\textrm{R}(t)$ are shown in Fig.~\ref{fig:HP_C_R_t}.
Below the critical point the behavior is similar to that observed for finite $N$. However, above the critical point, where the dynamics is linearized about only one of the two allowed semiclassical steady-state amplitudes, the rescaled concurrence is, as expected, quite different, owing to the more limited range of entangled states that the linearized (Gaussian) theory can accommodate.


\section{First-Order Phase Transition} \label{sect:af_model}

We now turn to the case of a fixed, positive interaction strength ($\lambda>0$) of the $\gamma=0$ model (Sec.~\ref{sub_sect:spin_model_3}) with variable effective field $h$. In the absence of dissipation this model exhibits two second-order transitions as $h$ is varied, one occurring at positive $h$ (the transition discussed in the previous section) and the other, equivalent transition occurring at negative $h$. However, in this section we show that with the addition of dissipation this model actually exhibits a first-order phase transition near $h\simeq 0$ (note that in the absence of dissipation no such transition exists). As in the previous section, we begin with a study of the linearized spin master equation, including an eigenvalue analysis and calculation of the probe transmission spectrum, after which we focus again on the entanglement properties of the system. For numerical calculations we will typically employ the set of normalized parameters $\{\lambda=1,\,\Gamma _a = 0.01,\,\Gamma_b = 0.2\}$, which correspond to a critical effective field strength $h_{\rm c}\simeq 0$.

\subsection{Linearized model} \label{sect:af_lin_model}

As before, we consider the thermodynamic limit and linearize the master equation~(\ref{eq:master_equation_gamma0_lmg_model}) about the mean-field state. To do so, we first find the semiclassical steady-state solutions and then expand the angular momentum operators around these mean-field solutions using the Holstein-Primakoff representation.

\subsubsection{Semiclassical steady-state solutions} \label{sect:af_semiclassical_equations}

From the (factorized) semiclassical equations of motion for $X$, $Y$, and $Z$, Eqs.~(\ref{eq:semicl(a)})-(\ref{eq:semicl(c)}) we again obtain the stable steady-state solutions. These exhibit discontinuities at the critical field strengths
\begin{eqnarray}
h_{\rm c} = \frac{1}{2} \left(\lambda- \sqrt{\lambda^2-\Gamma_b^2} \right) ,
\end{eqnarray}
and $h=0$. For $h<0$ the stable steady-state solutions are given by Eq.~(\ref{eq:semicl_ss_sols}), while for $h_{\rm c}<h <(\lambda+\sqrt{\lambda^2-\Gamma_b^2})/2$~\cite{SecondOrderFieldTransition} the stable steady states are given by Eqs.~(\ref{eq:semicl_ss_sols_(a)})-(\ref{eq:semicl_ss_sols_(c)}). While outside the region $0<h<h_{\rm c}$ the stable steady states are unique, inside the region $0<h<h_{\rm c}$ both steady-state solutions~(\ref{eq:semicl_ss_sols}) and~(\ref{eq:semicl_ss_sols_(a)})-(\ref{eq:semicl_ss_sols_(c)}) are in fact stable. However, for the characteristic parameters we consider here this region is very small ($h_{\rm c} \simeq 0.01$). Moreover, we have verified (from a linearized analysis) that the steady-state solution~(\ref{eq:semicl_ss_sols}) is more stable in the region $0<h<h_{\rm c}$ and thus we will only consider this solution in that region. Note that for larger values of the dissipation, $\Gamma_b$, this region becomes more pronounced (in this case all stable steady states should be considered~\cite{DissipativeTransition}), but this is beyond the regime we wish to consider here.

\begin{figure}[h!t]
\centerline{\includegraphics[width=8.6cm]{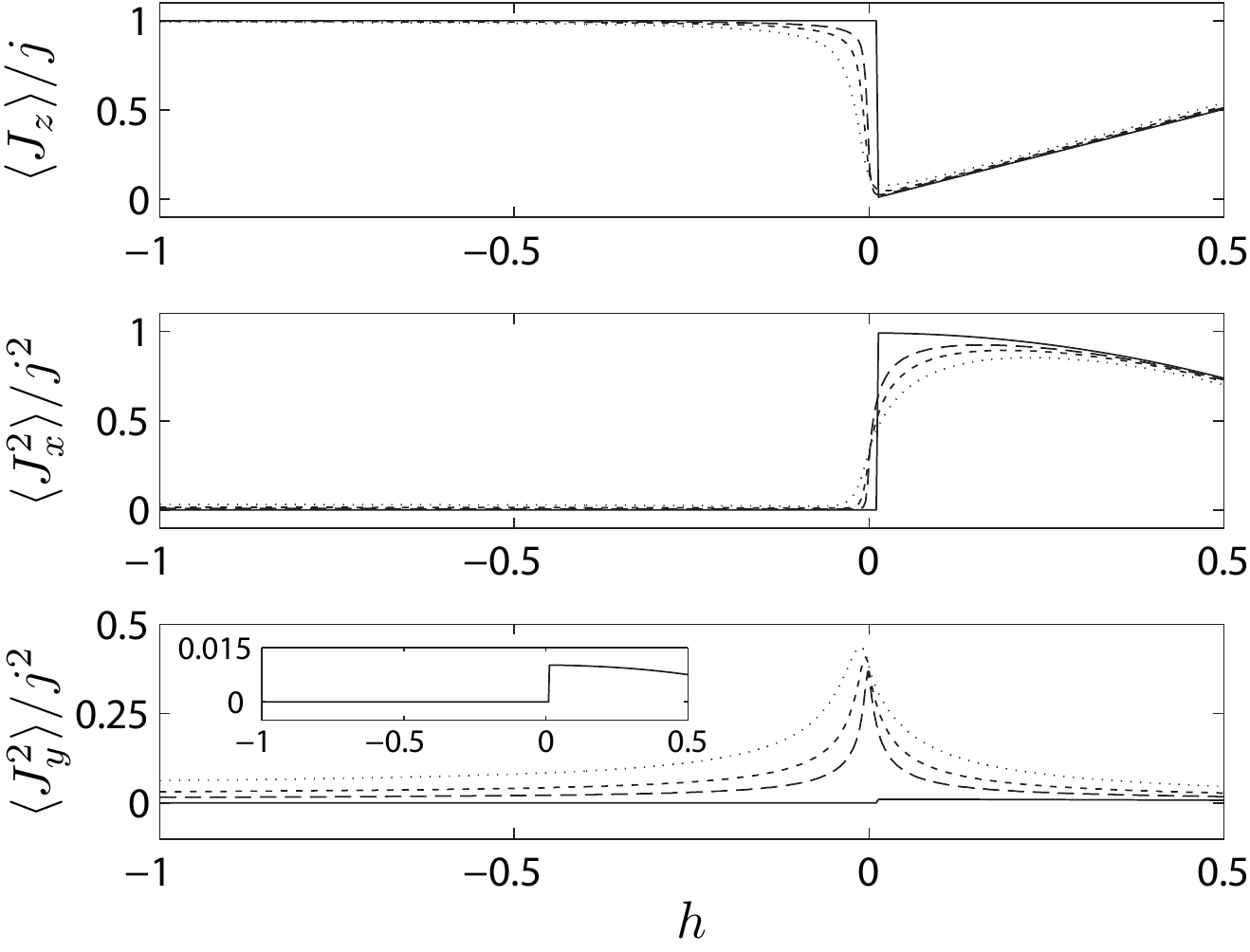}}
\caption{Semiclassical (solid line) and finite-$N$ steady-state solutions for $\lambda=1$, $\Gamma_a = 0.01$, $\Gamma_b=0.2$, and $N = 25$ (dotted), $50$ (short dashed line), $100$ (long dashed line). Note the inset in the bottom panel is a magnified plot of the semiclassical solution of $\av{J_y^2}/j^2$.} \label{fig:semicl_af}
\end{figure}

The relevant stable steady-state solutions $Z_\textrm{ss}$, $X_\textrm{ss}$ and $Y_\textrm{ss}$ are plotted in Fig.~\ref{fig:semicl_af}, together with results from numerical solutions of the master equation for a range of values of $N$ up to 100 (at which agreement between the two approaches is already quite good). The discontinuous jump of $Z_\textrm{ss}$ at $h_{\rm c}\simeq 0$ signifies the first-order phase transition. Note that for the case $\lambda<0$ the same first order transition occurs except that it is shifted to $-h_{\rm c}$ (i.e., in Fig.~\ref{fig:semicl_af} all curves are flipped about $h=0$).

\subsubsection{Holstein-Primakoff representation} \label{sect:af_hp_rep}

Here, we again include the quantum fluctuations for $N\gg1$ as a first-order correction by linearizing the spin operators around the semiclassical steady state via the HP representation. For $h<h_{\rm c}$ (normal phase) the linearized master equation is identical to Eq.~(\ref{eq:linearised_master_equation}) with $k=<$, while for $h>h_{\rm c}$ (broken phase) the linearized master equation is also identical to Eq.~(\ref{eq:linearised_master_equation}) but with $k=>$.

\subsubsection{Eigenvalue analysis} \label{sect:af_eigenvalues}

\begin{figure}[h!t]
\centerline{\includegraphics[width=8.6cm]{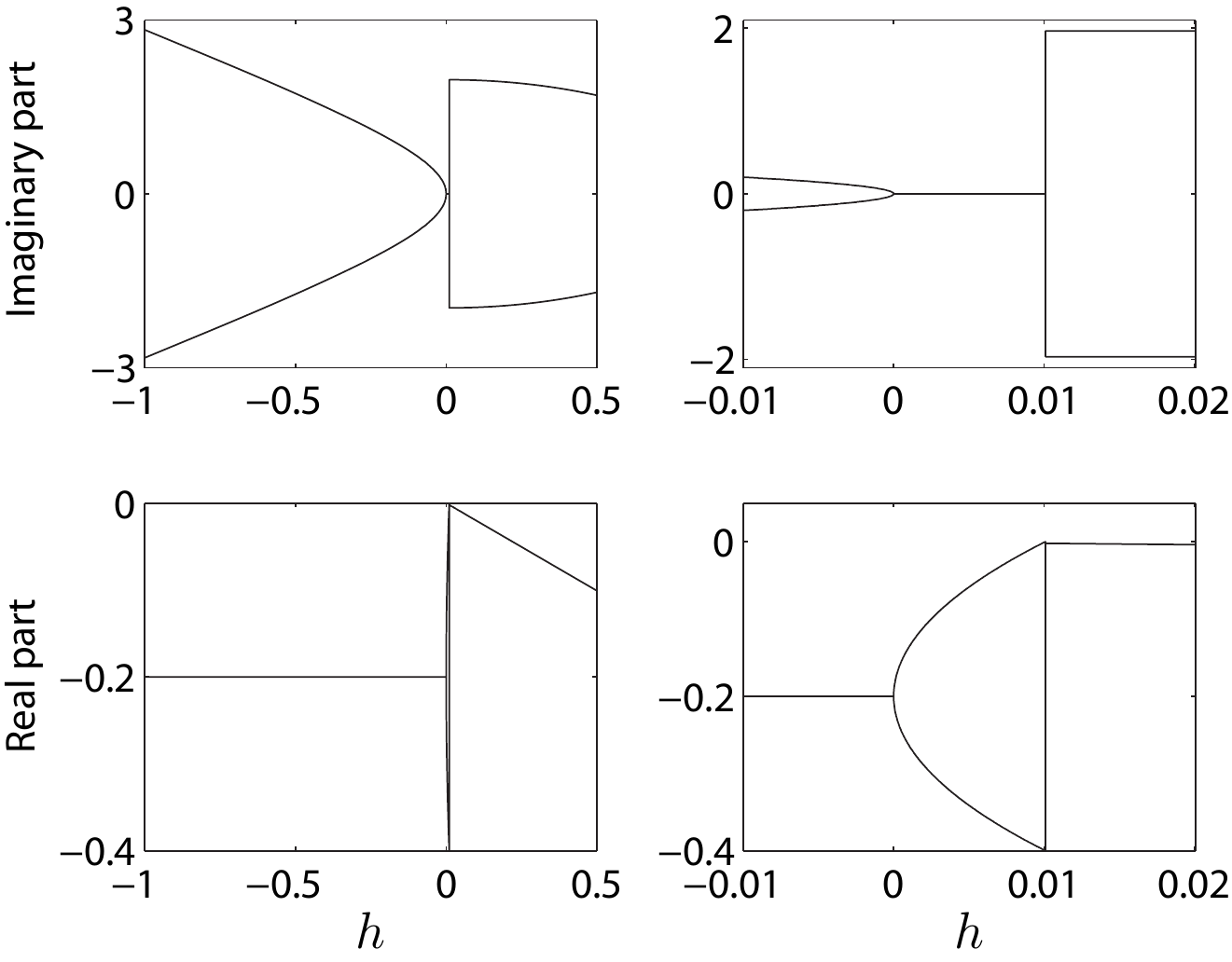}}
\caption{Eigenvalues of the linearized equations of motion, $\mu_\pm$, as given by Eqs.~(\ref{eq:eig_normal_phase}) and~(\ref{eq:eig_broken_phase}), for $\lambda=1$, and $\Gamma_b=0.2$. The right-hand column gives a magnified view of the region around $h_{\rm c}=0.0101$.} \label{fig:eig_lin_first}
\end{figure}

The eigenvalues of the linearized system, i.e., of the matrix $\mathbf{M}$, where $\dot{\vec{u}} = \mathbf{M} \vec{u}$ and $\vec{u} \equiv (\av{c}, \av{c^\dagger})^T$, for $h<h_{\rm c}$ are given by Eq.~(\ref{eq:eig_normal_phase}) while for $h>h_{\rm c}$ they are given by Eq.~(\ref{eq:eig_broken_phase}). In Fig.~\ref{fig:eig_lin_first} the real and imaginary parts of the eigenvalues are plotted for our characteristic set of parameters. In the normal phase ($h<h_{\rm c}$) we see that the imaginary parts go to zero at the point $h'=0 <h_{\rm c}$. In the region $h'<h<h_{\rm c}$ both eigenvalues are real and distinct, with one going to zero at $h_{\rm c}$ and the other going to $-2\Gamma_b$. This behavior is the same as that found for the second-order transition of the earlier section. However, immediately above the transition, $h>h_{\rm c}$, the eigenvalues become complex conjugate pairs with a nonzero real part that diminishes for $h \gg h_{\rm c}$ like $-\Gamma_b h/\lambda$. This discontinuous jump of the eigenvalues is an additional signature of the first-order phase transition.

\subsubsection{Transmission spectrum} \label{sect:af_transmission_spectrum}

We determine the probe transmission spectrum in the linearized regime following exactly the same calculations as outlined in Sec.~\ref{sect:transmission_spectrum}. The linearized Hamiltonian describing the full atom-cavity system is easily obtained; for $h<h_{\rm c}$ it is given by Eq.~(\ref{eq:linearised_atom_cavity_Hamiltonian}) with $\theta =0$ and $\phi=0$, while for $h>h_{\rm c}$ it is also given by Eq.~(\ref{eq:linearised_atom_cavity_Hamiltonian}), but with $\theta$ and $\phi$ given according to the semiclassical solutions Eqs.~(\ref{eq:semicl_ss_sols_(a)})-(\ref{eq:semicl_ss_sols_(c)}) as explained in Sec.~\ref{sect:hp_rep}.

\begin{figure}[h!t]
\centerline{\includegraphics[width=8.6cm]{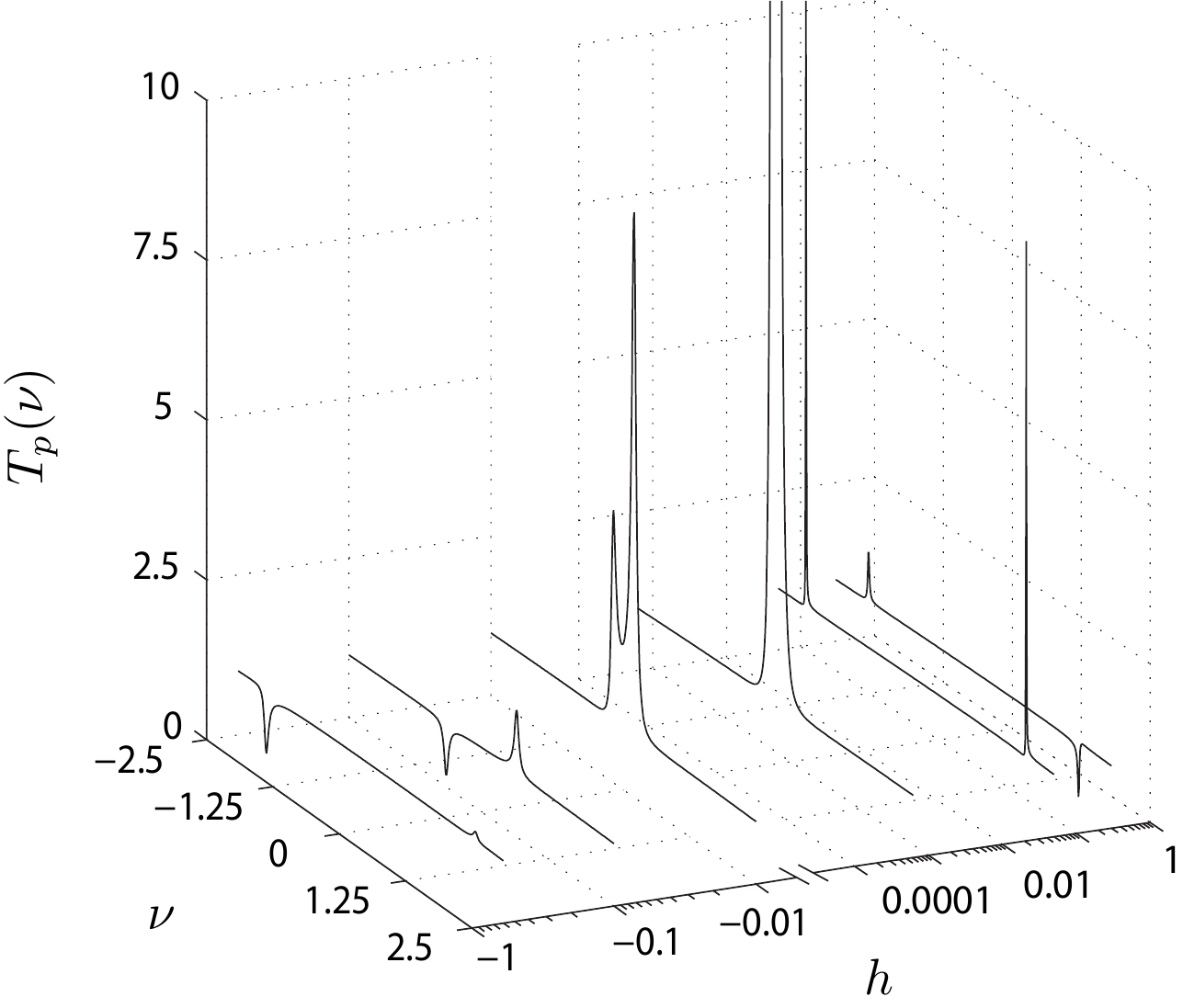}}
\caption{Transmission spectra in the linearized regime, for $h = -0.6, -0.1, -0.01, 6.25\times 10^{-4} (=h_{\rm c}), 0.05, 0.3$, with microscopic parameters $\kappa_a =0.3$, $\delta_a = 15$, $\lambda_a=2.7$, $\lambda_b=0.87$, and $\kappa_b=15$, giving $\lambda=1$, $\Gamma_a=0.01$ and $\Gamma_b = 0.05$.} \label{fig:trans_spect_af}
\end{figure}

Restricting ourselves again to a frequency range where $|\nu|\ll |\delta_a|,\kappa_b$ (also with $\kappa_a \ll |\delta_a|$), then as previously an approximate expression for the transmitted probe intensity can be derived in the normal phase for $h<h_{\rm c}$ and takes exactly the same form as Eq.~(\ref{eq:Tp_nu}).

In Fig.~\ref{fig:trans_spect_af} we plot the transmission spectrum for a series of values of $h$ across the critical point $h_{\rm c}$. In the normal phase ($h < h_{\rm c}$), we observe the same behavior as in the normal phase of the system in the previous section ($\lambda < \lambda_{\rm c}$), except that the orientations of the peaks and dips have inverted in accordance with the change of sign of the field ($h<0$). The central peak diverges as the critical point is approached from below in the normal phase, again signifying the phenomenon of critical slowing down in the vicinity of the phase transition. However, immediately above the critical point, the spectrum splits discontinuously into two sharp peaks of width $\sim\Gamma_bh/\lambda$, located at frequencies $\nu\simeq \pm2\lambda$. This jump from a single divergent peak at $\nu=0$ to a two-peaked spectrum offers a pronounced, observable signature of the first-order transition.

\subsection{Entanglement} \label{sect:af_ent}

\subsubsection{Steady-state entanglement} \label{sect:steady_state_entanglement_af}

\begin{figure}[h!t]
\centerline{\includegraphics[width=8.6cm]{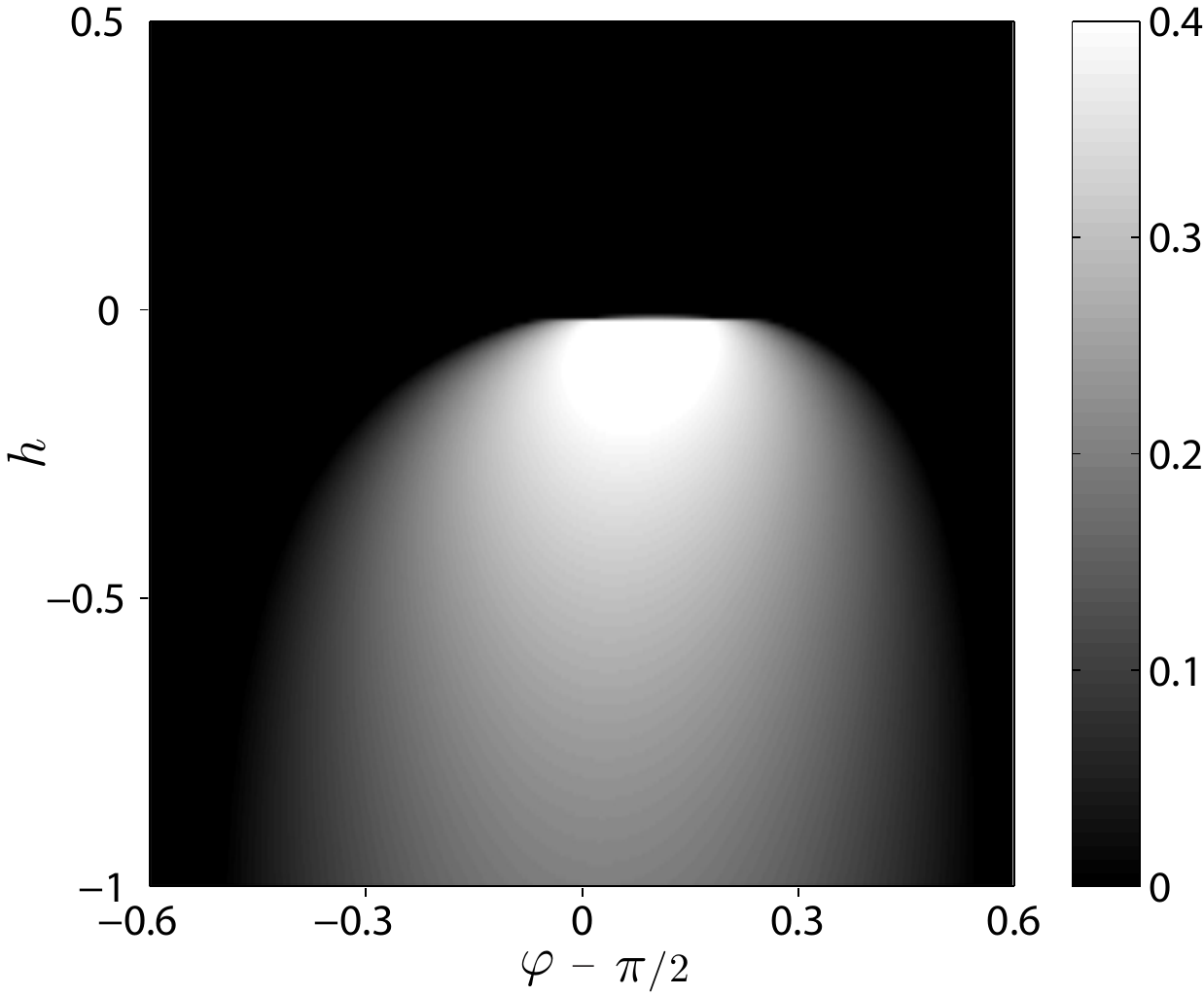}}
\caption{Entanglement measure $\textrm{max}\{ 0,C_\varphi\}$ as a function of $h$ and $\varphi$ for $N=100$, $\lambda=1$, $\Gamma_a = 0.01$, and $\Gamma_b=0.2$.} \label{fig:fN_C_varphi_af}
\end{figure}

We compute, as before, the entanglement measures $C_\varphi$ and $C_\textrm{R}$, both numerically for finite $N$ and analytically for $N\gg1$ in the linearized regime. Fig.~\ref{fig:fN_C_varphi_af} shows a plot of $C_\varphi$ as a function of $h$ and $\varphi$ for $N=100$. We see that, well away from the critical point, substantial entanglement is present over a broad range of angles $\varphi$. As the critical point is approached from below, significant entanglement persists, but for a somewhat narrower range of angles $\varphi$. However, immediately above the critical point the entanglement drops suddenly to zero for all values of $\varphi$.

\begin{figure*}[h!t]
\centerline{\includegraphics[width=14cm]{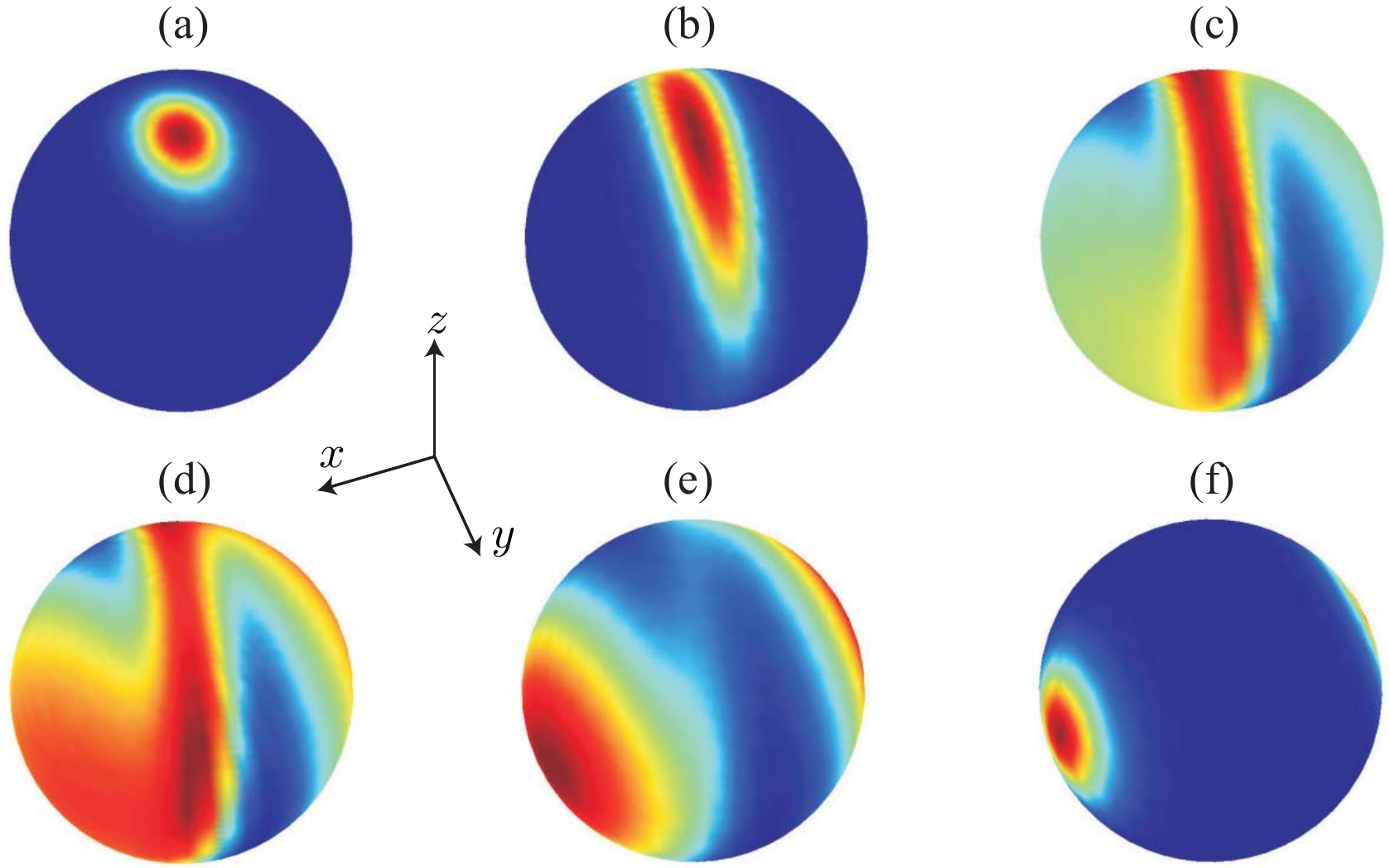}}
\caption{(Color online) Steady-state spin $Q$-function, $Q_\textrm{s}(\eta)$, on the Bloch sphere for (a) $h = -0.5$, (b) $h=-0.01$, (c) $h=2.5 \times 10^{-3}$, (d) $h= 5 \times 10^{-3}$, (e) $h=0.015 $, and (f) $h=0.15$, with $N=50$, $\lambda=1$, $\Gamma_a = 0.01$, and $\Gamma_b=0.2$. Note that dark blue corresponds to the minimum value of zero of $Q_{\rm s}(\eta)$ while dark red indicates the maximum value of $Q_{\rm s}(\eta)$.} \label{fig:spin_qfunc_first}
\end{figure*}

To help understand these results we again utilize the atomic coherent state representation and study the spin $Q$-function $Q_{\rm s}(\eta)$. In Fig.~\ref{fig:spin_qfunc_first} we plot $Q_{\rm s}(\eta)$ on the Bloch sphere for a series of values of $h$ in the vicinity of the first-order transition. Well below the critical point, in the normal phase, $Q_{\rm s}(\eta)$ is a single peaked function with little angular dependence. Correspondingly, $C_\varphi$ is nonzero over a broad range of $\varphi$, with a maximum close to $\varphi=\pi/2$ (i.e., near $C_x$). Again, note that this slight shift of the optimum away from $\varphi=\pi/2$ is a consequence of the dissipation ($\Gamma_b$) in the system.

As $h$ increases towards the critical point, $Q_{\rm s}(\eta)$ becomes increasingly stretched along the $y$ axis. As the critical point is traversed $Q_{\rm s}(\eta)$ rapidly rotates around from the $y$ axis towards the $x$ axis, and splits into the familiar two-lobed structure associated with the two semiclassical steady-state amplitudes of the broken phase. At the same time as the critical point is approached, the range of $\varphi$ over which $C_\varphi$ remains finite narrows and immediately above the critical point it drops abruptly to zero for all choices of $\varphi$. This behavior is akin to the behavior we observed for large interaction strength in the regime of the previous section, where $\av{J_x^2}$ becomes of order $j^2=N^2/4$ (see Fig.~\ref{fig:semicl_af}), which severely restricts the range of $\varphi$ for which $C_\varphi>0$. Note that at larger values of $h$ than displayed in Fig.~\ref{fig:fN_C_varphi_af}, the entanglement, $C_\varphi$, once again becomes nonzero (centered around $\varphi \approx 0$) coinciding with the broken phase behavior of the second-order transition discussed in the previous section.

\begin{figure}[h!t]
\centerline{\includegraphics[width=8.6cm]{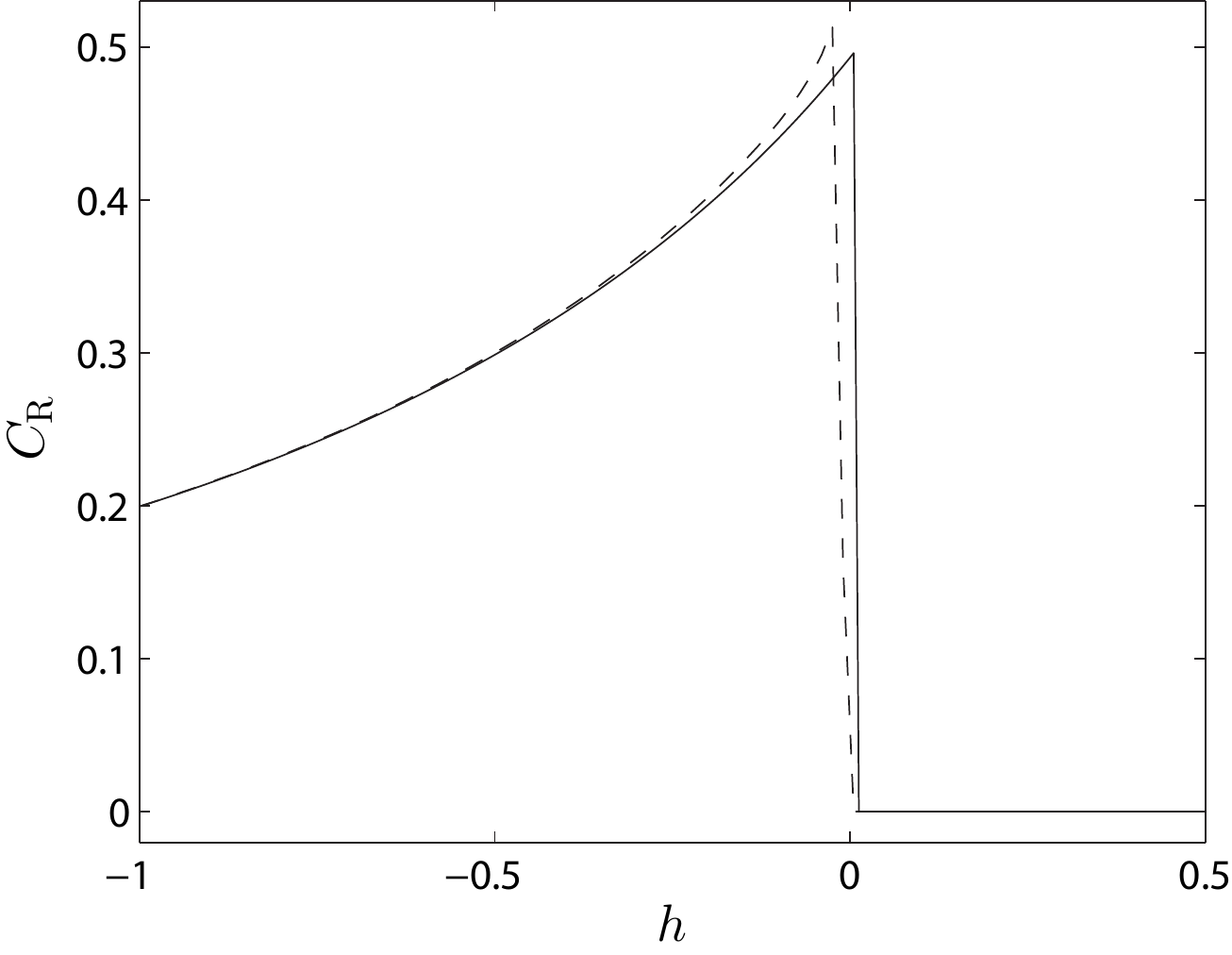}}
\caption{Rescaled concurrence $C_\textrm{R}$ versus $h$ for $N=100$ (dashed line) and in the thermodynamic limit (solid line) with $\lambda=1$, $\Gamma_a = 0.01$, and $\Gamma_b = 0.2$.} \label{fig:C_R_first_fN_and_HP_combo}
\end{figure}

In Fig.~\ref{fig:C_R_first_fN_and_HP_combo} we plot the rescaled concurrence $C_\textrm{R}$ as a function of the effective field strength $h$ and again find that close to the critical point, $h_{\rm c}$, the entanglement reaches its peak value. Although the equivalent closed system would not feature a maximum in the entanglement near $h_{\rm c}$ (due to the complete absence of a phase transition), this result is in agreement with a conjecture concerning entanglement in open systems at quantum critical points~\cite{Schneider}.

In the linearized treatment ($N\gg 1$) we obtain very similar plots of $C_\varphi$ to those of finite $N$ (Fig.~\ref{fig:fN_C_varphi_af}) and for $C_\textrm{R}$ the result is shown in Fig.~\ref{fig:C_R_first_fN_and_HP_combo}. In the limit where we consider $\Gamma_a\simeq 0$ we can again obtain an approximate expression for the rescaled concurrence (for $h\leq h_{\rm c}$) given, in this instance, by
\begin{eqnarray}
C_\textrm{R}^\textrm{HP} &\simeq& \frac{\lambda(\sqrt{(h-\Lambda/2)(h-h_{\rm c})+\lambda^2}-\lambda)}{4(h-\Lambda/2)(h-h_{\rm c})}
\nonumber
\\
&\simeq & \frac{1}{2} - \frac{1}{2}  \frac{h_{\rm c}-h}{\lambda} ~~~ \textrm{for} ~~~ h_{\rm c}- h\ll\lambda .
\end{eqnarray}
This again has a maximum value of $0.5$ at the critical point, and, for large $|h|$, drops off like $1/|h|$, in reasonable agreement with the plots.

\subsubsection{Entanglement dynamics} \label{sect:entanglement_dynamics_af}

\begin{figure}[h!t]
\centerline{\includegraphics[width=8.6cm]{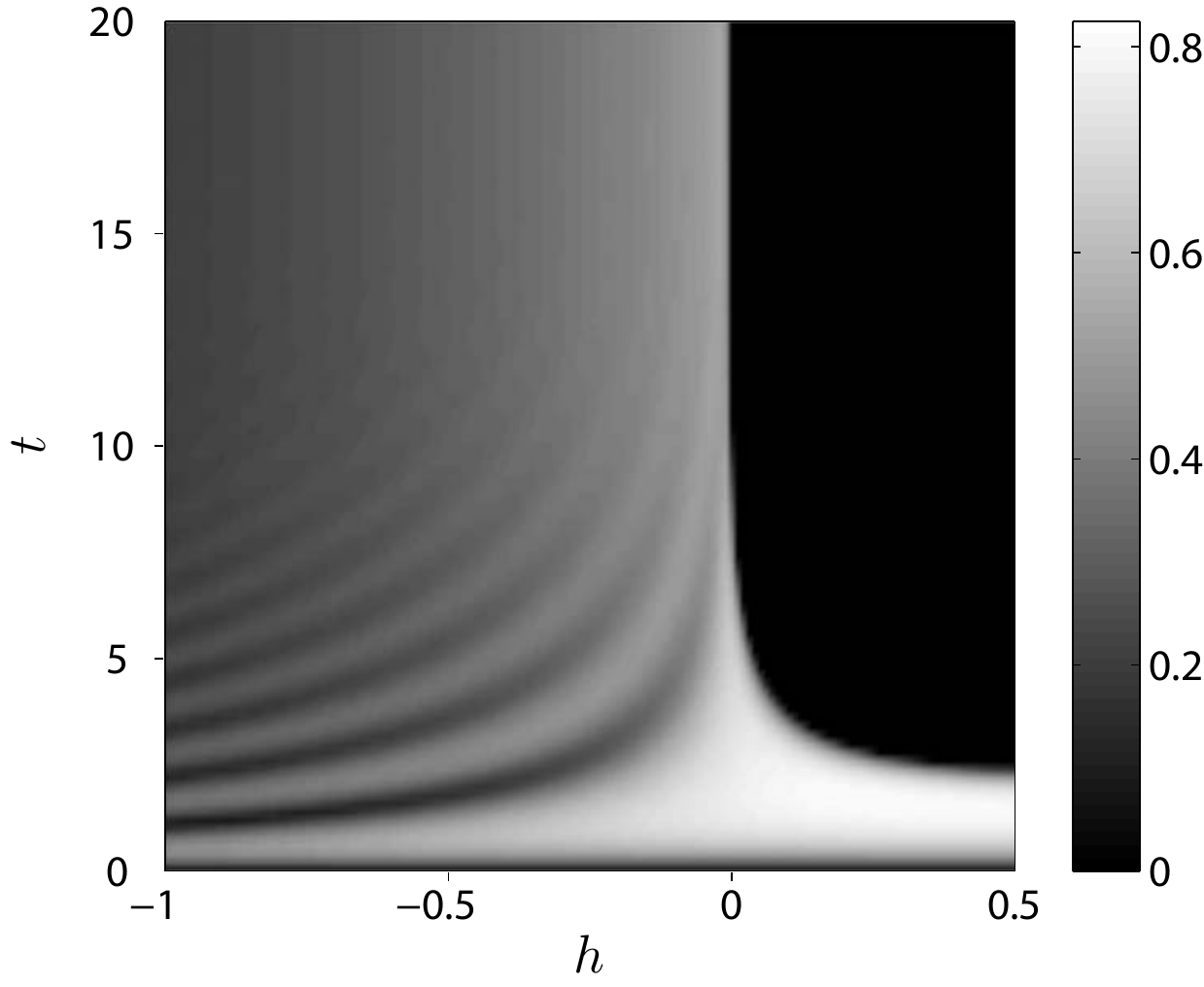}}
\caption{Rescaled concurrence $C_\textrm{R}(t)$ for $N=100$, with $\lambda=1$, $\Gamma_a = 0.01$, and $\Gamma_b=0.2$.} \label{fig:fN_C_R_t_af}
\end{figure}

Finally, in Fig.~\ref{fig:fN_C_R_t_af} for $N=100$  we illustrate the time-dependent behavior of the rescaled concurrence,  $C_\textrm{R}(t)$, for varying $h$, given an initial (unentangled) state with all spins up. Once again, we observe an interesting oscillatory behavior of $C_\textrm{R}(t)$, with, in particular, highly entangled states generated by the Hamiltonian dynamics at short times (for almost all values of $h$), before dissipation has had time to play a significant role. For the linearized regime ($N\gg1$) a similar plot of  $C_\textrm{R}(t)$ can be obtained which agrees well with the finite $N$ result for $h<h_{\rm c}$ but shows zero entanglement for almost all values of $h>h_{\textrm c}$ because of the restricted linearization around only one of the two permitted semiclassical steady-state amplitudes.

\section{Conclusions}
\label{sect:conclusion}

We have proposed in this paper a feasible cavity QED setup, consisting of a collective atomic pseudospin and two quantized cavity modes, which realizes a dissipative version of the LMG model in which the interacting spin system displays both first- and second-order nonequilibrium quantum phase transitions. The lossy cavity's output light fields can be utilized to monitor the system as the model parameters are varied; specifically, we showed that the transmission spectra vary dramatically in the vicinity of the transition, with features that are characteristic of the criticality. A further important result is the steady-state entanglement criticality at the QPT and the possibility of directly observing this via homodyne detection of the cavity output fields. In particular, the entanglement can be quantified rather directly in terms of measurable atomic quadrature variances. We also observed an important sensitivity of the entanglement measure to the quadrature phase angle in the critical regimes, which we were able to interpret by employing an atomic phase space distribution. Finally, we have considered how entanglement evolves in this system, observing not only the criticality at the QPT at long times (corresponding to the steady state), but also a rich transient behavior at shorter times.

For future studies, it is clear that the system we have proposed offers a variety of opportunities, such as (i) investigating phase transitions in response to variation of the strength of dissipation (i.e., $\Gamma_b$), (ii) examining a system of multiple (separately addressable) atomic pseudospins all coupled to the same quantized cavity modes, which would permit the study of entanglement between different spin blocks~\cite{EntLMGBlock}, (iii) controlled preparation of robust (insensitive to noise/environment), highly entangled states by evolution from an initial product state~\cite{BECMicheli,MolmerGHZ}, (iv) measurement of more general atomic spin correlations and their evolution with time, which can also provide signatures of criticality in QPT's~\cite{Das06}, (v) extending our system to accommodate more complex spin models, e.g., by adding additional lasers to the setup explained in Sec.~\ref{sect:general_model} to realize the so-called ``two-field model''~\cite{EntLMGConcurrReview}, and (vi) imposing some spatial variation on the cavity mode to provide, for example, short ranged interactions, which could be uniform or quasirandom.

\begin{acknowledgments}
The authors thank A. Daley and H. Carmichael for discussions and acknowledge support from the Austrian Science Foundation and from the Marsden Fund of the Royal
Society of New Zealand.
\end{acknowledgments}


\appendix

\section{Coefficients of the Atom-Cavity Hamiltonian in the Linearized Regime} \label{sect:appone}

In Sec.~\ref{sect:transmission_spectrum} we gave the general form of the linearized Hamiltonian of the joint atom-cavity system, Eq.~(\ref{eq:linearised_atom_cavity_Hamiltonian}). The coefficients of this Hamiltonian in terms of the system parameters, $h,\lambda_a,\lambda_b,\Gamma_b$ and the angles $\theta,\phi$ from Sec.~\ref{sect:hp_rep} are
\begin{eqnarray}
\delta_c & = & 2h\cos{\theta} +2 \sin{\theta} \left[2 \lambda X_\textrm{ss} \cos{\phi} \right.
\nonumber
\\
&& ~~~~~~ \left. - \Gamma_b(Y_\textrm{ss}\cos{\phi}-X_\textrm{ss}\sin{\phi}) \right], \\
A & = & \frac{\lambda_a}{2} \left[ (1+\cos{\theta})+(1-\cos{\theta})(\sin{\phi}+i\cos{\phi})^2\right], \nonumber \\ \\
B_1 & = & \frac{\lambda_b}{2}\left[ (1-\cos{\theta})(\sin{\phi}+i\cos{\phi})^2\right] , \\
B_2 & = & \frac{\lambda_b}{2}(1+\cos{\theta}).
\end{eqnarray}
Note that for $\lambda<\lambda_{\rm c}$ one has $\theta=0$ and $\phi=0$, giving the simplified expressions $\delta_c = 2h$, $A=\lambda_a$, $B_1=0$, and $B_2=\lambda_b$. Similar to Sec.~\ref{sect:hp_rep}, we can also derive simplified expressions in the limit $\lambda \gg \lambda_{\rm c}$, i.e., for $\lambda \rightarrow \infty$, one has $\delta_c = 4\lambda_a$, $A=0$, $B_1 =-\lambda_b/2$, and $B_2 =\lambda_b/2$



\begin{thebibliography}{99}

\bibitem{Jaksch05}
For recent reviews, see
D. Jaksch and P. Zoller, Ann. Phys. (N.Y.) {\bf315}, 52 (2005), and
I. Bloch, Nature Phys. {\bf1}, 23 (2005).

\bibitem{Greiner02}
M. Greiner, O. Mandel, T. Esslinger, T. H\"ansch, and I. Bloch, Nature {\bf415}, 39 (2002).

\bibitem{Micheli06}
A. Micheli, G.~K. Brennen, and P. Zoller, Nature Phys. {\bf2}, 341 (2006).

\bibitem{Porras04}
D. Porras and J.~I. Cirac, Phys. Rev. Lett. {\bf92}, 207901 (2004).

\bibitem{originalLMG123}
H.~J. Lipkin, N. Meshkov, and A.~J. Glick, Nucl. Phys.  {\bf62}, 188, 199, 211 (1965);
N. Meshkov, A.~J. Glick, and H.~J. Lipkin, {\it ibid}. {\bf62}, 199 (1965);
A.~J. Glick, H.~J. Lipkin, and N. Meshkov, {\it ibid}. {\bf62}, 211 (1965).

\bibitem{Osterloh02}
A. Osterloh {\em et al}., Nature {\bf416}, 608 (2002).

\bibitem{Osborne02}
T.~J. Osborne and M.~A. Nielsen, Phys. Rev. A {\bf66}, 032110 (2002).

\bibitem{GVidal03}
G. Vidal, J.I. Latorre, E. Rico, and A. Kitaev, Phys. Rev. Lett. {\bf90}, 227902 (2003).

\bibitem{EntLMGSecondOrder}
J. Vidal. G. Palacios, and R. Mosseri, Phys. Rev. A {\bf69}, 022107 (2004).

\bibitem{EntLMGFirstOrder}
J. Vidal. R. Mosseri, and J. Dukelsky, Phys. Rev. A {\bf69}, 054101 (2004).

\bibitem{EntLMGDynamics}
J. Vidal. G. Palacios, and C. Aslangul, Phys. Rev. A {\bf70}, 062304 (2004).

\bibitem{EntLMGEntropy}
J. I. Latorre, R. Or\'us, E. Rico, and J. Vidal, Phys. Rev. A {\bf71}, 064101 (2005).

\bibitem{EntLMGCUT}
S. Dusuel and J. Vidal, Phys. Rev. B {\bf71}, 224420 (2005).

\bibitem{EntLMGBlock}
T. Barthel, S. Dusuel, and J. Vidal, Phys. Rev. Lett. {\bf97}, 220402 (2006).

\bibitem{EntLMGConcurrReview}
J. Vidal, Phys. Rev. A {\bf73}, 062318 (2006).


\bibitem{BECLMGSemiclassical}
G.~J. Milburn, J. Corney, E.~M. Wright, and D.~F. Walls, Phys. Rev. A {\bf55}, 4318 (1997).

\bibitem{BECMicheli}
A. Micheli, D. Jaksch, J.~I. Cirac, and P. Zoller, Phys. Rev. A {\bf67}, 013607 (2003).

\bibitem{MolmerGHZ}
K. M{\o}lmer and A. S{\o}rensen, Phys. Rev. Lett. {\bf82}, 1835 (1999).

\bibitem{Fleischhauer1}
R.~G. Unanyan, M. Fleischhauer, N.~V. Vitanov, and K. Bergmann, Phys. Rev. A {\bf66}, 042101 (2002);
R.~G. Unanyan and M. Fleischhauer, Phys. Rev. Lett. {\bf90}, 133601 (2003).


\bibitem{Berman94}
P. Berman, ed., {\it Cavity Quantum Electrodynamics} (Academic Press, Boston, 1994).


\bibitem{Hepp73}
K. Hepp and E.~H. Lieb, Ann. Phys. (N.Y.) {\bf76}, 360 (1973);
Phys. Rev. A {\bf8}, 2517 (1973).

\bibitem{Wang73}
Y.~K. Wang and F.~T. Hioe, Phys. Rev. A {\bf7}, 831 (1973).

\bibitem{Hioe73}
F.T. Hioe, Phys. Rev. A {\bf8}, 1440 (1973).

\bibitem{Carmichael73}
H.~J. Carmichael, C.~W. Gardiner, and D.~F. Walls, Phys. Lett. A {\bf46}, 47 (1973).

\bibitem{Duncan74}
G. Cromer Duncan, Phys. Rev. A {\bf9}, 418 (1974).


\bibitem{Dicke}
R.~H. Dicke, Phys. Rev. {\bf93}, 99 (1954).


\bibitem{Emary03a}
C. Emary and T. Brandes, Phys. Rev. Lett. {\bf90},  044101 (2003).

\bibitem{Emary03b}
C. Emary and T. Brandes, Phys. Rev. E {\bf67},  066203 (2003).


\bibitem{Lambert04}
N. Lambert, C. Emary, and T. Brandes, Phys. Rev. Lett. {\bf92},  073602 (2004).

\bibitem{Lambert05}
N. Lambert, C. Emary, and T. Brandes, Phys. Rev. A {\bf71},  053804 (2005).

\bibitem{Reslen05}
J. Reslen, L. Quiroga, and N.~F. Johnson, Europhys. Lett. {\bf69}, 8 (2005).


\bibitem{Rzaznewski75}
K. Rza\.{z}ewski, K. W\'odkiewicz, and W. \.{Z}acowicz, Phys. Rev. Lett. {\bf35}, 432 (1975).

\bibitem{Dimer07}
F. Dimer, B. Estienne, A.~S. Parkins, and H.~J. Carmichael, Phys. Rev. A {\bf75}, 013804 (2007).


\bibitem{Bonifacio76}
R. Bonifacio and L.~A. Lugiato, Opt. Commun. {\bf19}, 172 (1976); Phys. Rev. Lett. {\bf40}, 1023 (1978);
Phys. Rev. A {\bf18}, 1129 (1978).

\bibitem{Drummond78}
P.~D. Drummond and H.~J. Carmichael, Opt. Commun. {\bf27}, 160 (1978).

\bibitem{Walls78}
D.~F. Walls, P.~D. Drummond, S.~S. Hassan, and H.~J. Carmichael, Prog. Theor. Phys. Suppl. {\bf64}, 307 (1978).

\bibitem{Drummond80}
P.~D. Drummond, Phys. Rev. A {\bf22}, 1179 (1980).

\bibitem{Carmichael80}
H.~J. Carmichael, J. Phys. B {\bf13}, 3551 (1980).


\bibitem{Schneider}
S. Schneider and G.~J. Milburn, Phys. Rev. A {\bf65}, 042107 (2002).


\bibitem{QuantumNoise}
C.~W. Gardiner and P. Zoller, \emph{Quantum Noise} (Springer-Verlag, Berlin, 1992).

\bibitem{CollettGardiner84_85}
M.~J. Collett and C.~W. Gardiner, Phys. Rev. A {\bf30}, 1386 (1984);
C.~W. Gardiner and M.~J. Collett, {\em ibid}. {\bf31}, 3761 (1985).


\bibitem{WisemanQuadrature}
H.~M. Wiseman and G.~J. Milburn, Phys. Rev. A {\bf47}, 642 (1993).


\bibitem{QOToolbox}
For numerical solutions of the atomic collective-spin master equation we make use of:
S.~M. Tan, {\em Quantum Optics and Computation Toolbox for Matlab}, available at http://www.qo.phy.auckland.ac.nz/qotoolbox.html.

\bibitem{Holstein40}
T. Holstein and H. Primakoff, Phys. Rev. {\bf58}, 1098 (1940).

\bibitem{Ressayre75}
E. Ressayre and A. Tallet, Phys. Rev. A {\bf11}, 981 (1975).

\bibitem{vonCube06}
Ch. von Cube, S. Slama, M. Kohler, C. Zimmermann, and Ph.~W. Courteille,
Fortschr. Phys. {\bf54}, 726 (2006).

\bibitem{Klinner06}
J. Klinner, M. Lindholdt, B. Nagorny, and A. Hemmerich, Phys. Rev. Lett. {\bf96}, 023002 (2006).

\bibitem{SecondOrderTransition}
Note that for the case $h<0$ an analogous second-order transition occurs at $-\lambda_{\rm c}$.

\bibitem{WallsandMilburn}
D. F. Walls and G. J. Milburn, \emph{Quantum Optics} (Springer-Verlag, Berlin, 1994).


\bibitem{EntanglementCriteria}
J.~K. Korbicz, J.~I. Cirac, and M. Lewenstein, Phys. Rev. Lett. {\bf95}, 120502 (2005);
{\it ibid}. {\bf95}, 259901 (2005).

\bibitem{OrigConcurrece}
W.~K. Wooters, Quant. Inf. Comp. {\bf1}, 27 (2001).

\bibitem{MolmerConcurrSymm}
X. Wang and K. M{\o}lmer, Eur. Phys. J. D {\bf18}, 385 (2002).

\bibitem{DissipativeTransition}
S. Morrison and A.~S. Parkins (unpublished)

\bibitem{SecondOrderFieldTransition}
Note that at $(\lambda+\sqrt{\lambda^2-\Gamma_b^2})/2$ a second order phase transition analogous to the one already presented in Sec.~\ref{sect:ferro_model} occurs, and thus for the first order transition we will focus on $h\ll (\lambda+\sqrt{\lambda^2-\Gamma_b^2})/2$.

\bibitem{Das06}
A. Das, K. Sengupta, D. Sen, and B.~K. Chakrabarti, Phys. Rev. B {\bf74}, 144423 (2006).

\end{thebibliography}
\end{document}